# KMT-2021-BLG-0171Lb and KMT-2021-BLG-1689Lb: Two Microlensing Planets in the KMTNet High-cadence Fields with Followup Observations

Hongjing Yang (杨弘靖)[1]⋆ , Weicheng Zang (臧伟呈)[1] , Andrew Gould[3,4], Jennifer C. Yee[5] ,
Kyu-Ha Hwang[6], Grant Christie[7], Takahiro Sumi[8], Jiyuan Zhang (张纪元)[1] , Shude Mao (毛淑德)[1,2]
    (Leading Authors)
Michael D. Albrow[9], Sun-Ju Chung[6,10], Cheongho Han[11], Youn Kil Jung[6], Yoon-Hyun Ryu[6],
In-Gu Shin[5,11], Yossi Shvartzvald[12], Sang-Mok Cha[6,13], Dong-Jin Kim[6], Hyoun-Woo Kim[6],
Seung-Lee Kim[6,10], Chung-Uk Lee[6], Dong-Joo Lee[6], Yongseok Lee[6,13], Byeong-Gon Park[6,10],
Richard W. Pogge[4,6]
    (The KMTNet Collaboration)
John Drummond[14], Dan Maoz[15], Jennie McCormick[16], Tim Natusch[7,17], Matthew T. Penny[18] ,
Wei Zhu (祝伟)[1]
    (The MAP & μFUN Follow-up Teams)
Ian A. Bond[19], Fumio Abe[20], Richard Barry[21], David P. Bennett[21,22], Aparna Bhattacharya[21,22],
Martin Donachie[23], Hirosane Fujii[8], Akihiko Fukui[24,25], Yuki Hirao[8], Yoshitaka Itow[8,20],
Rintaro Kirikawa[8], Iona Kondo[8], Naoki Koshimoto[26,27], Man Cheung Alex Li[23], Yutaka Matsubara[20],
Yasushi Muraki[20], Shota Miyazaki[8], Greg Olmschenk[21], Clément Ranc[21], Nicholas J. Rattenbury[23],
Yuki Satoh[8], Hikaru Shoji[8], Stela Ishitani Silva[28,21], Daisuke Suzuki[29], Yuzuru Tanaka[8],
Paul J. Tristram[30], Tsubasa Yamawaki[8], Atsunori Yonehara[31]
    (The MOA Collaboration)

[1] *Department of Astronomy, Tsinghua University, Beijing 100084, China*
[2] *National Astronomical Observatories, Chinese Academy of Sciences, Beijing 100101, China*
[3] *Max-Planck-Institute for Astronomy, Königstuhl 17, 69117 Heidelberg, Germany*
[4] *Department of Astronomy, Ohio State University, 140 W. 18th Ave., Columbus, OH 43210, USA*
[5] *Center for Astrophysics | Harvard & Smithsonian 60 Garden St., Cambridge, MA 02138, USA*
[6] *Korea Astronomy and Space Science Institute, Daejon 34055, Republic of Korea*
[7] *Auckland Observatory, Auckland, New Zealand*
[8] *Department of Earth and Space Science, Graduate School of Science, Osaka University, Toyonaka, Osaka 560-0043, Japan*
[9] *University of Canterbury, Department of Physics and Astronomy, Private Bag 4800, Christchurch 8020, New Zealand*
[10] *University of Science and Technology, Korea, (UST), 217 Gajeong-ro Yuseong-gu, Daejeon 34113, Republic of Korea*
[11] *Department of Physics, Chungbuk National University, Cheongju 28644, Republic of Korea*
[12] *Department of Particle Physics and Astrophysics, Weizmann Institute of Science, Rehovot 76100, Israel*
[13] *School of Space Research, Kyung Hee University, Yongin, Kyeonggi 17104, Republic of Korea*
[14] *Possum Observatory, Patutahi, Gisborne, New Zealand*
[15] *School of Physics and Astronomy, Tel-Aviv University, Tel-Aviv 6997801, Israel*
[16] *Farm Cove Observatory, Centre for Backyard Astrophysics, Pakuranga, Auckland, New Zealand*
[17] *Institute for Radio Astronomy and Space Research (IRASR), AUT University, Auckland, New Zealand*
[18] *Department of Physics and Astronomy, Louisiana State University, Baton Rouge, LA 70803 USA*
[19] *Institute of Natural and Mathematical Sciences, Massey University, Auckland 0745, New Zealand*
[20] *Institute for Space-Earth Environmental Research, Nagoya University, Nagoya 464-8601, Japan*
[21] *Code 667, NASA Goddard Space Flight Center, Greenbelt, MD 20771, USA*
[22] *Department of Astronomy, University of Maryland, College Park, MD 20742, USA*
[23] *Department of Physics, University of Auckland, Private Bag 92019, Auckland, New Zealand*
[24] *Department of Earth and Planetary Science, Graduate School of Science, The University of Tokyo, 7-3-1 Hongo, Bunkyo-ku, Tokyo 113-0033, Japan*
[25] *Instituto de Astrofísica de Canarias, Vía Láctea s/n, E-38205 La Laguna, Tenerife, Spain*
[26] *Department of Astronomy, Graduate School of Science, The University of Tokyo, 7-3-1 Hongo, Bunkyo-ku, Tokyo 113-0033, Japan*
[27] *National Astronomical Observatory of Japan, 2-21-1 Osawa, Mitaka, Tokyo 181-8588, Japan*
[28] *Department of Physics, The Catholic University of America, Washington, DC 20064, USA*
[29] *Institute of Space and Astronautical Science, Japan Aerospace Exploration Agency, 3-1-1 Yoshinodai, Chuo, Sagamihara, Kanagawa, 252-5210, Japan*
[30] *University of Canterbury Mt. John Observatory, P.O. Box 56, Lake Tekapo 8770, New Zealand*
[31] *Department of Physics, Faculty of Science, Kyoto Sangyo University, 603-8555 Kyoto, Japan*

**ABSTRACT**
Follow-up observations of high-magnification gravitational microlensing events can fully exploit their intrinsic sensitivity to detect extrasolar planets, especially those with small mass ratios. To make followup more uniform and efficient, we develop a system, HighMagFinder, based on the real-time data from the Korean Microlensing Telescope Network (KMTNet) to automatically alert possible ongoing high-magnification events. We started a new phase of follow-up observations with the help of HighMagFinder in 2021. Here we report the discovery of two planets in high-magnification microlensing events, KMT-2021-BLG-0171 and KMT-2021-BLG-1689, which were identified by the HighMagFinder. We find that both events suffer the "central-resonant" caustic degeneracy. The planet-host mass-ratio is $q \sim 4.7 \times 10^{-5}$ or $q \sim 2.2 \times 10^{-5}$ for KMT-2021-BLG-0171, and $q \sim 2.5 \times 10^{-4}$ or $q \sim 1.8 \times 10^{-4}$ for KMT-2021-BLG-1689. Together with two events reported by Ryu et al. (2022), four cases that suffer such degeneracy have been discovered in the 2021 season alone, indicating that the degenerate solutions may have been missed in some previous studies. We also propose a new factor for weighting the probability of each solution from the phase-space. The resonant interpretations for the two events are disfavored under this consideration. This factor can be included in future statistical studies to weight degenerate solutions.

## 1 INTRODUCTION

With more than 120[1] detected planets, gravitational microlensing has proven to be a powerful method for probing extrasolar planets (Mao & Paczynski 1991; Gould & Loeb 1992). Unlike other methods, microlensing can discover wide-orbit and small planets around all types of stellar objects.

The typical rate of microlensing events toward the Galactic bulge is $\sim 10^{-5} - 10^{-6}$ per monitored star per year (e.g., Sumi et al. 2013; Mróz et al. 2019). Therefore, detecting microlensing events, and consequently planetary events, requires wide-area surveys that monitor a large number of stars.

The light curves of most microlensing events are symmetric and bell-shaped (Paczyński 1986). The typical Einstein timescales $t_E$ of such light curves are $\sim 20$ days. Planetary signals are usually small perturbations on the light curves. The half duration of the planetary perturbation (Gould & Loeb 1992) is approximately

$$t_p \sim t_E \sqrt{q} \sim 1.5(q/10^{-5})^{1/2} \text{ hr}, \quad (1)$$

where $q$ is the planet to host mass ratio. Assuming that at least 10 points are needed to claim a detection, the observational cadence should be $\gtrsim 3$ hr$^{-1}$ to detect $q \sim 10^{-5}$ planets (e.g., an Earth-mass planet around a 0.3 $M_\odot$ host). These microlensing planets are critical for building a statistical sample to extend the mass range to Earth-mass planets.

For many years, many microlensing planets were discovered by a combination of wide-area low-cadence surveys to find microlensing events and intensive follow-up observations to capture the planetary perturbations (Gould & Loeb 1992). Another strategy of finding microlensing planets, pioneered by the OGLE and MOA projects, is conducting wide-area, high-cadence surveys toward the Galactic bulge. The Korea Microlensing Telescope Network (KMTNet, Kim et al. 2016) aims at this strategy. KMTNet continuously monitors a broad area at relatively high-cadence toward the Galactic bulge from three 1.6 m telescopes equipped with 4 deg$^2$ FOV cameras at the Cerro Tololo Inter-American Observatory (CTIO) in Chile (KMTC), the South African Astronomical Observatory (SAAO) in South Africa (KMTS), and the Siding Spring Observatory (SSO) in Australia (KMTA). Since 2016, KMTNet has monitored a total of (3, 7, 11, 2) fields at cadences of $\Gamma \sim (4, 1, 0.4, 0.2)$ hr$^{-1}$ (See Figure 12 of Kim et al. 2018). In the majority of the fields, the cadence is too low to reliably detect most $q \sim 10^{-5}$ planets, thus follow-up observations are needed.

The cadence of the KMTNet prime fields, $\Gamma \geqslant 4$ hr$^{-1}$, can potentially detect $q \sim 10^{-5}$ planets alone, e.g., OGLE-2019-BLG-1053 (Zang et al. 2021). However, usually the ideal cadence cannot be achieved in reality because of (i) time gaps between observatories, and (ii) bad weather conditions at one or more sites. These issues cause planetary signals to be missed or the confidence of planetary detections to be lowered. To resolve these issues and fully extract the potential of microlensing events, followup observations are still needed for the KMTNet high cadence survey fields. Therefore, we perform a followup program for all KMTNet survey fields. We focus on high-magnification events that are intrinsically more sensitive to planets (Griest & Safizadeh 1998).

Meanwhile, with the growing number of discovered microlensing events each year by the KMTNet ($\sim 3000$), $> 200$ events at any given time must be tracked to determine whether they require followup observations, because high-magnification events vary quickly and the magnifications of ongoing events are difficult to predict. In addition, it is difficult to create a uniform statistical sample from a sample of high-magnification events selected by eye[2].

Therefore, we developed HighMagFinder, a system to automatically monitor all ongoing events based on the KMTNet real-time data. Every three hours, it alerts possible high-magnification events to the observers. The system helped us to discover six new planets in 2021 with much less ($< 10\%$) manpower compared to previous followup efforts.

In this paper, we begin by describing the HighMagFinder algorithm in Section 2. Then, we report the detection of planets in two high-magnification microlensing events, KMT-2021-BLG-0171 and KMT-2021-BLG-1689. Both of these events were identified by the HighMagFinder. In Section 3, we introduce the observations of these events including both survey and followup data. We then report the light curve modelling of the two events in Section 4, the properties of the microlens sources in Section 5, and the physical parameters of the planetary systems in Section 6. Finally, we discuss the role of followup observations in 7.1 and the newly discovered degeneracy for high-magnification events in 7.2. We estimate the phase-space factor for the degenerate solutions in 7.3.

---

* E-mail: yang-hj19@mails.tsinghua.edu.cn
[1] https://exoplanetarchive.ipac.caltech.edu, as of 2022 Jan. 23.

[2] Although not impossible, see Gould et al. (2010).

## 2 HIGHMAGFINDER

In 2019, once KMTNet started alerting events from all fields, it became more difficult to identify potential high-magnification events by eye from the huge number of ongoing events. We develop HighMagFinder to automatically fit and classify all events based on the KMTNet real-time pipeline data[3].

The HighMagFinder is scheduled to run at the same cadence as KMTNet updates real-time data (every 3 hours) and reports all the possible high magnification events. Here we describe the algorithm of the HighMagFinder.

Limited by the telescope resources for follow-up, we focus on events with maximum (intrinsic) magnification $A_{\max} > 50$, which corresponds to the microlens impact parameter $u_0 < 0.02$. The algorithms below are designed and optimized to find such events with the fewest false negatives. The thresholds are mostly selected by experience and can be altered if the criteria for interesting targets change.

For each event, we first remove data points with large FWHM and sky background to create a cleaner light curve and lower the false positive rate. The threshold of FWHM is taken to be FWHM < 9.0, 6.5, 7.0 for KMTA, KMTC, and KMTS, respectively. The sky background upper limit for all sites is set to be 3000. However, all data points within ±5 days around the peak are protected. Based on experience with the KMTNet data, we then rescale the errorbars of all data points by a factor of 2.0, 1.5, 1.6 for KMTA, KMTC, and KMTS, respectively.

Secondly, a series of point-source point-lens (PSPL, Paczyński 1986) microlensing models are used to fit the cleaned light curve. The model consists of three parameters, $(t_0, u_0, t_E)$, where $t_0$ is the time when the source is closest to the center of lens mass, $u_0$ is the impact parameter in the unit of Einstein radius $\theta_E$, and $t_E$ is the Einstein radius crossing time or microlensing timescale. We start with fitting the light curve with all PSPL parameters set free, and the result is the best-fit model with $\chi^2_{\text{best}}$. Then we perform three additional fits, where $u_0$ is fixed to 0.01, 0.025, and 0.05, respectively. The chi-square's of these fits are $\chi^2_{u_0=0.01}$, $\chi^2_{u_0=0.025}$, and $\chi^2_{u_0=0.05}$. We also fit the light curve with a flat line, and the resulting chi-square is $\chi^2_{\text{flat}}$.

By comparing the goodness of these fits, we can estimate the possibility of an event to become high-magnification and decide whether or not to alert it. Events that satisfy the following conditions are alerted as possible high magnification events.

$$-5 \text{ day} < t_{\text{now}} - t_{0,\text{best}} < 3 \text{ day}; \tag{2}$$

$$\chi^2_{u_0=0.05} - \chi^2_{u_0=0.025} > 0.3; \tag{3}$$

$$u_{0,\text{best}} < 0.025 \quad \text{or} \quad \chi^2_{u_0=0.025} - \chi^2_{\text{best}} < 7; \tag{4}$$

$$t_{0,u_0=0.01} - t_{0,\text{best}} < 10 \text{ day}; \tag{5}$$

$$\chi^2_{\text{flat}} - \chi^2_{\text{best}} > 1000. \tag{6}$$

Eq. 2 selects events that are in a close time window because events that peak in the far future are uncertain. Eqs. 3 and 4 select events that could reach a high magnification. The fixed $u_0$ values, as well as the thresholds, can be adjusted as the observational strategy varies, e.g., if the observers focus on higher or lower magnification events.

[3] https://kmtnet.kasi.re.kr/~ulens/

Eqs. 5 and 6 prevent false positives caused by low signal-to-noise ratio light curves. In principle, these criteria should be different for fields with different cadences. But using the values optimized for the the lowest cadence fields, $\Gamma = 0.2$ hr$^{-1}$, are also satisfactory for any higher cadence fields.

Finally, for possible high magnification events, we generate a report page. The report contains a table which lists all the fitting parameters and a figure for each event. See Figure 1 as an example. On the figure, the left panel shows the full light curve and the right panel shows a zoom-in plot near the peak. All the key parameters are labelled on the figure. After each run, observers will receive the report and manually check it. For the true positives, observers will then decide on the follow-up strategy based on the fitting results.

The formal operation of HighMagFinder started on 2021-06-08. During its operation in the 2021 season, the HighMagFinder alerted 352 events (on average ∼3 new alerts every day), and about 1/3 of them turned out to be real $A_{\max} > 50$ high-magnification events. The majority of false positives are caused by the uncertainty before the peak. Fig. 2 shows the cumulative distributions of all alerted events and the true positives. Both of the distributions are uniform for $u_0 < 0.014$ or $A_{\max} > 70$, which implies the selection criteria do not create bias for these events. For comparison, Gould et al. (2010) were only able to achieve such homogeneity for $A_{\max} > 200$ using by-eye methods. Thus, a homogeneous statistical sample of followup events can be selected by HighMagFinder. In addition, the false negative rate of HighMagFinder is < 2%.

By following up, six new planets have been discovered in five events identified by this system, they are KMT-2021-BLG-0171Lb (this work), KMT-2021-BLG-0247Lb (in prep.), KMT-2021-BLG-1547Lb (in prep.), KMT-2021-BLG-1689Lb (this work), and an event with two planets KMT-2021-BLG-0185Lb,c (Han et al. 2022, in prep.).

## 3 OBSERVATION

### 3.1 Preamble

Here we report two planets in events that were identified as high-magnification by the HighMagFinder. Although the HighMagFinder did not start official operations until June 2021, KMT-2021-BLG-0171 was identified by HighMagFinder on 2021-04-19 in its trial run. KMT-2021-BLG-1689 was alerted by HighMagFinder during its regular operations on 2021-07-12, UT 12:30 (JD' ∼ 9408.0). Below we give a detailed observation history of these events.

### 3.2 Surveys

Both events are located in the Galactic bulge. The coordinates are listed in Table 1.

The increase of the source brightness in both events was first found by the AlertFinder algorithm (Kim et al. 2018) of the KMTNet survey. KMT-2021-BLG-0171 was alerted on 2021-03-29, UT 04:51 (HJD' ≡ HJD − 2450000 ∼ 9302.7) and KMT-2021-BLG-1689 was alerted on 2021-07-12, UT 04:31 (HJD' ∼ 9407.7).

The Microlensing Observations in Astrophysics (MOA, Bond et al. 2001; Sumi et al. 2003) group, utilizing the 1.8 m telescope of the Mt. John Observatory in New Zealand, independently found KMT-2021-BLG-1689 one day after the KMTNet's discovery and marked it as MOA-2021-BLG-258. Hereafter we use the name KMT-2021-BLG-1689 following the first discovery.

The images from the KMTNet survey were mainly acquired in the

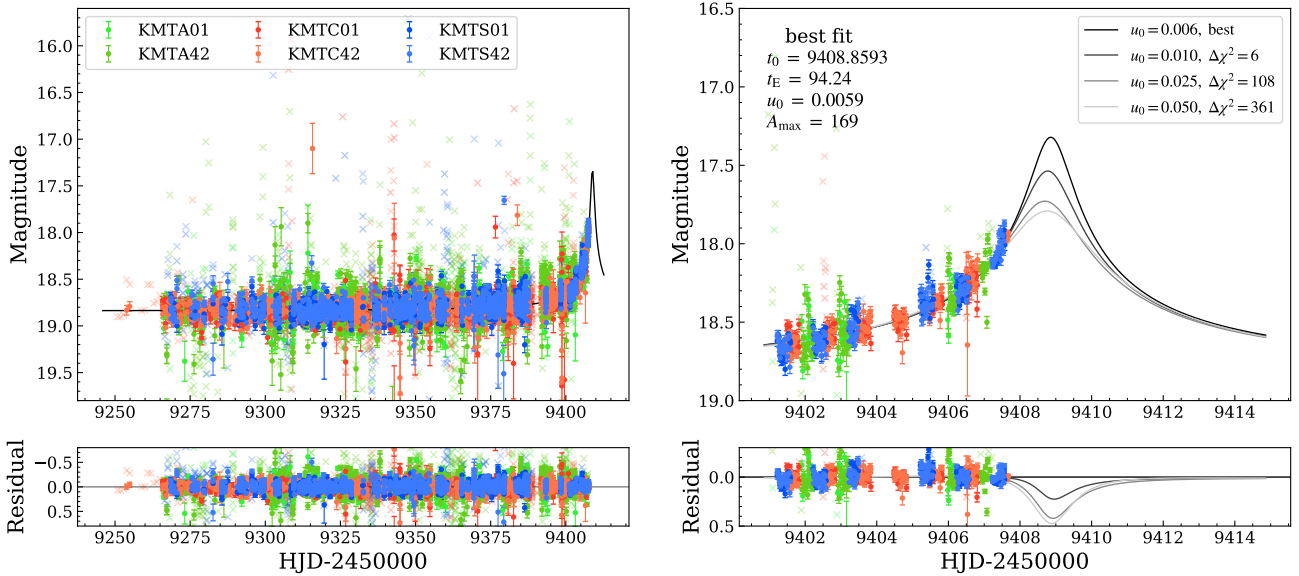

**Figure 1.** The HighMagFinder report figure of KMT-2021-BLG-1689 at 1.5 days before the event reached its peak. The left panel shows the full light curve from both the observational data and the best-fit model. The excluded data are marked as "x". The right panel shows a zoom-in plot near the peak together with the different models. The best-fit parameters and the $\Delta\chi^2$ of each fit are labeled.

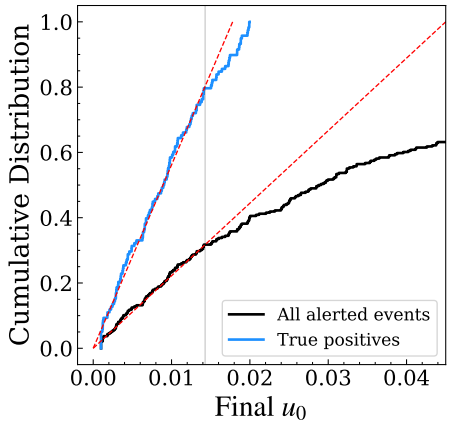

**Figure 2.** The cumulative distribution of events alerted by HighMagFinder during its 2021 season regular operation. The black and blue curves represent all alerted events and the true positives, respectively. The red lines indicate uniform distributions. For $u_0 < 0.014$ or $A_{\max} > 70$, the two distributions are uniform. For comparison, see Figure 1 of Gould et al. (2010).

*I*-band, and a fraction of images were obtained in the *V*-band for measuring the color. The images from the MOA survey were mainly taken in the MOA-red wide band, which is the sum of the standard Cousins *R*- and *I*-bands, and a fraction of images were taken in *V*-band.

### 3.3 Followup

At UT 07:02 on 2021-04-20 HJD′ ∼ 9324.8, W.Zang found by eye that KMT-2021-BLG-0171 could be a high-magnification candidate and sent an alert to the Microlensing Astronomy Probe (MAP) and $\mu$FUN Follow-up Team and scheduled high-cadence followup observations with Las Cumbres Observatory (LCO) global network. $\mu$Fun observations were taken by Possum Observatory (Possum) and Farm Cove Observatory (FCO) in New Zealand. The LCO global network took observations from its 1.0m telescopes located at SSO (LCOA), CTIO (LCOC), and SAAO (LCOS), with the SDSS-$r'$ filter. Possum and FCO respectively took observations using their 0.36m telescopes without a filter.

KMT-2021-BLG-1689 was alerted by HighMagFinder in 2021-07-12, UT 12:30 (JD′ ∼ 9408.0). Because the peak of this event was predicted to be $I \gtrsim 16$ mag during the New Zealand and Australia zone, which is faint for most $\mu$FUN sites there, the MAP & $\mu$FUN Follow-up Team only sent an alert to Auckland Observatory (AO) at UT 07:37 on 2021-07-13 (JD′ ∼ 9408.8). High-cadence follow-up observations were immediately taken by the 0.4m telescope at AO with a Wratten #12 filter. Moreover, there are no LCO follow-up data for this event due to the limited time allocated in 2021 July.

The data used in the light-curve analysis were reduced using the various difference image analysis (Tomaney & Crotts 1996; Alard & Lupton 1998) pipelines: pySIS (Albrow et al. 2009) for the KMTNet data and $\mu$Fun (Possum, FCO, AO) data, Bond et al. (2001) for the MOA data and ISIS (Alard & Lupton 1998; Alard 2000; Zang et al. 2018) for the LCO data. For the KMTC01 data in both events, we conduct pyDIA[4] photometry to measure the source color.

## 4 LIGHT-CURVE ANALYSIS

### 4.1 Preamble

Figures 3 and 4 show the observed data together with the best fit models for KMT-2021-BLG-0171 and KMT-2021-BLG-1689, respectively. The light curves for both of these two events deviate from the PSPL light curve by a bump near the peak. The bump of KMT-2021-BLG-0171 is captured by KMTA and LCOA observations, and the bump of KMT-2021-BLG-1689 is captured by AO and MOA observations. These sorts of anomalies can be produced by either caustic crossing or cusp approaching in a binary-lens (2L1S) event (Mao & Paczynski 1991; Gould 1992), or the second source in a binary-source (1L2S) event (Gaudi 1998). Therefore, we perform both 2L1S and 1L2S analyses for these two events.

---

[4] MichaelDAlbrow/pyDIA: Initial Release on Github, doi:10.5281/zenodo.268049

**Table 1.** Event Names and Locations

| Event | $(\alpha, \delta)_{J2000}$ | $(l, b)$ | Field | KMTNet cadence |
|---|---|---|---|---|
| KMT-2021-BLG-0171 | (17:56:58.18,−30:05:34.58) | (0.267°, −2.714°) | KMT01, KMT42 | 4 hr$^{-1}$ |
| KMT-2021-BLG-1689 | (17:58:18.62,−30:08:43.12) | (0.366°, −2.991°) | KMT01, KMT42, MOA-GB8 | 4 hr$^{-1}$ |

A standard 2L1S model requires seven parameters to describe the magnification $A(t)$. The first three are the same as PSPL ($t_0$, $u_0$, $t_E$), where the $u_0$ is measured relative to the angular Einstein radius $\theta_E$ of the total lens mass. The next three ($q$, $s$, $\alpha$) define the binary geometry: the binary mass ratio, the projected separation between the binary components normalized to the Einstein radius, and the angle between the source trajectory and the binary axis in the lens plane. The last parameter is the source radius normalized by the Einstein radius, $\rho = \theta_*/\theta_E$, where $\theta_*$ is the angular radius of the source star. In addition, for each data set $i$, two flux parameters ($f_{S,i}$, $f_{B,i}$) represent the flux of the source star and the blend flux. The observed flux, $f_i(t)$, is calculated from

$$f_i(t) = f_{S,i} A(t) + f_{B,i}. \quad (7)$$

For each event, we generally start with locating the local $\chi^2$ minima by searching over a grid of parameters ($\log s$, $\log q$, $\log \rho$, $\alpha$). The grid consists of 61 equally spaced values in $-1.5 \leqslant \log s \leqslant 1.5$, 56 equally spaced values in $-5.5 \leqslant \log q \leqslant 0$, 9 equally spaced values in $-4.0 \leqslant \log \rho \leqslant -1.6$, and 20 equally spaced values in $0° \leqslant \alpha < 360°$. For each set of initial parameters, we fix $\log q$, $\log s$ and $\log \rho$, and allow $t_0$, $u_0$, $t_E$, $\alpha$ to vary. In each grid, we find the minimum $\chi^2$ by Markov Chain Monte Carlo (MCMC) using the emcee ensemble sampler (Foreman-Mackey et al. 2013). After finding one or more local minima on the ($\log s, \log q, \log \rho$) space, each local is further refined by allowing all seven parameters to vary in an MCMC.

For the standard 1L2S model, the light curve is the superposition of two 1L1S curves. There are at least eight parameters (Hwang et al. 2013): ($t_{0,1}, u_{0,1}, \rho_1$) and ($t_{0,2}, u_{0,2}, \rho_2$) describe the impact time, impact parameter and the size of two sources, respectively. The Einstein radius crossing time $t_E$ is the same for the two sources. Finally, the flux ratio of two sources is $q_f = f_{S,1}/f_{S,2}$. The flux ratio of two sources might differ in different bands, so if the event is observed in multiple bands, a separate $q_f$ should be used for each band.

For both 1L2S and 2L1S models, we further examine the microlens parallax effect which is caused by the orbital motion of Earth (Gould 1992, 2000, 2004). The microlens-parallax is

$$\boldsymbol{\pi}_E = \frac{\pi_{rel}}{\theta_E} \frac{\boldsymbol{\mu}_{rel}}{\mu_{rel}}, \quad (8)$$

where ($\pi_{rel}, \boldsymbol{\mu}_{rel}$) are the lens-source relative parallax and proper motion.

In addition, if the finite-source effect appears in the light curve, i.e., the source crosses the caustic curves, the limb-darkening effect should be included. We use the linear limb-darkening law,

$$S_\lambda(\mu) = S_\lambda(1) \left[1 - u_\lambda(1 - \mu)\right], \quad (9)$$

where $S_\lambda(1)$ is the surface brightness at the center of the source, $\mu$ is the cosine of the angle between the normal to the stellar surface and the line of sight, and $u_\lambda$ is the limb-darkening coefficient at wavelength $\lambda$. For each event, the limb-darkening coefficients are inferred from the effective temperature $T_{eff}$ (Claret & Bloemen 2011), which is estimated in Section 5.

The detailed light curve analyses for the two events are presented separately below.

### 4.2 KMT-2021-BLG-0171

#### 4.2.1 Binary-lens (2L1S) modelling

We conduct an initial grid search for 2L1S solutions as described in Section 4.1. The upper panel in Figure 5 shows the $\chi^2$ distribution in the projected ($\log s, \log q$) plane from the initial grid search, which indicates the distinct minima are within $-0.4 \leqslant \log s \leqslant 0.4$, $-5.0 \leqslant \log q \leqslant -3.2$ and $-3.1 \leqslant \log \rho \leqslant -2.5$. We therefore perform a denser grid search with this smaller parameter space which is shown in the lower-left panel in Figure 5. The second grid search reveals two distinct local minima, A and B. However, there are still unresolved features near $\log s \sim 0$, so we further conduct a refined grid search in $-0.03 \leqslant \log s \leqslant 0.03$ and $-5.0 \leqslant \log q \leqslant -4.0$. The result of the third grid search is shown in the lower-right panel, where two new local minima, C and D, are resolved.

We then investigate the best-fit model with all the standard 2L1S parameters set free using MCMC. Because in Models C and D the source star crossed the caustic, we include limb-darkening effect of the source. From Section 5, we infer the effective surface temperature of the source is $\sim 5200$ K and consequently the limb-darkening coefficients are $(u_I, u_r, u_R, u_V) = (0.5451, 0.6624, 0.6368, 0.7200)$ (Claret & Bloemen 2011). For the unfiltered data, we approximately take $u_U \sim (u_I + u_R)/2$.

The best-fit parameters are listed in Table 2. The modelling indicates that A is the best solution, however, B, C and D are disfavored by only $\Delta\chi^2 = 0.2$, 3.7, and 5.7. The model light curves together with the data are shown in Fig. 3. The caustic structures are shown in Figure 6. The (A, B) solutions are central cusp approaches, and the (C, D) solutions are resonant caustic crossings. We will further discuss the degeneracy between (A, B) and (C, D) in Section 7.2.

We further investigate the parallax effect. We notice that the parallax signal from KMTC42 baseline data is not consistent with all the other sites (fields), thus we exclude the data outside $t_0 \pm 50$ days of KMTC42. We fitted $u_0 > 0$ and $u_0 < 0$ scenarios for each solution to consider the ecliptic degeneracy (Skowron et al. 2011). In general, with two more parameters, the parallax fits only improve the $\chi^2$ by $\sim 8$ for all solutions. However, we find that the east component of the parallax vector $\pi_{E,E}$ is well constrained for all solutions, while the constraint on the north component $\pi_{E,N}$ is considerably weaker. See Figure 7. This is simply because the Earth's motion is roughly in the East direction. More precisely, the minor axis of the likelihood contour is aligned with the projected position of the Sun at $t_0$ (e.g., Gould et al. 1994; Smith et al. 2003). The best-fit parameters of each parallax model are shown in Table 3.

### 4.2.2 Binary-source (1L2S) modelling

We search for 1L2S solutions using MCMC, and the best fit model is disfavored by $\Delta\chi^2 \sim 11$ compared to the 2L1S A model (see Table 4). Such a small $\Delta\chi^2$ means the 1L2S model also describes the observed light curve reasonably well. However, this solution does not seem to be self-consistent. If we assume the two sources have similar $T_{\rm eff}$ (given that $q_{f,I} \approx q_{f,r}$), then the brightness of the source should be proportional to the square of the radius, $f_s \propto \rho^2$. From Table 4 we know $\rho_1 < 4.6 \times 10^{-3}$ and $\rho_2 \sim 1.5 \times 10^{-3}$, therefore, we expect

$$q_f = \frac{f_{s,2}}{f_{s,1}} \sim \left(\frac{\rho_2}{\rho_1}\right)^2 > \left(\frac{1.5 \times 10^{-3}}{4.6 \times 10^{-3}}\right)^2 \sim 0.1, \quad (10)$$

which is much larger than the modeled flux ratio $q_{f,I} = 0.0065 \pm 0.0009$.

To explore this conflict more quantitatively, we investigate the color effects. Gaudi (1998) proposed that the binary-source interpretation can be tested by the color difference of two sources with different luminosity. Thus, we employ an extra pySIS reduction for the KMTNet $V-$band images. We then refine the solution by MCMC with the new data included. The best-fit parameters are shown in Table 4. The $q_{f,V}$ is clearly measured because $V-$band data cover the anomaly region with three data points. This allow us to measure the color difference between two sources,

$$(V-I)_{s,2} - (V-I)_{s,1} = -2.5 \log\left(\frac{q_{f,V}}{q_{f,I}}\right), \quad (11)$$

and the $I-$magnitude difference,

$$I_{s,2} - I_{s,1} = -2.5 \log q_{f,I}. \quad (12)$$

The second source is marked in the color-magnitude diagram (CMD, left panel of Figure 10).

We immediately see that the two modelled sources have nearly the same color, which is what we would expect for an effect due to lensing of a single source. That is, if the anomaly were due to a binary source, one might expect that the sources would be different colors, especially given the large magnitude difference between them. By contrast, if the anomaly is due to a magnification effect, such as a binary lens, the source color should be constant throughout the event (apart from very small difference due to limb-darkening). Hence, the fact that the two sources have roughly the same color tends to support the 2L1S interpretation over the 1L2S interpretation.

This analysis also allows us to quantify the conflict between the source flux ratio and the source radius ratio. From the color difference, we infer the source angular radius ratio by Adams et al. (2018),

$$\log\left(\frac{\rho_2}{\rho_1}\right) = 0.378 \left[(V-I)_{s,2} - (V-I)_{s,1}\right] - 0.2 \left[I_{s,2} - I_{s,1}\right], \quad (13)$$

with a typical uncertainty of ~10%. We calculate the inferred $\rho_2/\rho_1$ for each MCMC chain, and compare it with the directly modeled $\rho_2/\rho_1$ in Fig. 11. The figure shows there are no solutions for which the inferred $\rho_2/\rho_1$ matches the value of $\rho_2/\rho_1$ from the fit. Therefore, we rule out the 1L2S interpretation of this event.

### 4.3 KMT-2021-BLG-1689

#### 4.3.1 Binary-lens (2L1S) modelling

As for the first event, we first locate the local $\chi^2$ minima by an initial grid search. The upper panel in Figure 8 shows the $\chi^2$ distribution in the projected ($\log s$, $\log q$) plane from the initial grid search. The result shows two distinct local minima E and F. Except for E and F, the majority of the (unresolved) local minima are located within $-0.25 \leqslant \log s \leqslant 0.25$, $-4.5 \leqslant \log q \leqslant -2.5$, and $-3.1 \leqslant \log \rho \leqslant -2.8$. We therefore perform two denser grid searches which are shown in the two lower panels in Figure 5. We adopt $\Delta \log q = 0.05$ for both new grid searches, using $\Delta \log s = 0.01$ and $\Delta \log s = 0.002$, respectively. For the $\rho$ values, the width of the anomaly bump is approximately the upper-limit of the source diameter, thus the light curve indicates that

$$\rho \leqslant \frac{\Delta t_{\rm anom}}{2 t_{\rm E}} \sim 0.0016, \quad \log \rho \lesssim 2.8, \quad (14)$$

where $\Delta t_{\rm anom}$ is the width of the anomaly signal. This is consistent with the result of the initial grid search. Thus we only adopt two values of $\log \rho = -2.8, -3.1$. The refined grid searches reveal six more distinct local minima in total, A, A′, B, B′, C and D. However, A (B) and A′ (B′) become the same solution if we allow $\rho$ to be a free parameter. In total, we resolved six local minima labeled from A to F in Figure 8.

We then investigate the best-fit model with all the standard 2L1S parameters set free using MCMC. We infer the effective surface temperature of the source to be ~ 4600 K from Section 5, and consequently the limb-darkening coefficients to be $(u_I, u_R, u_V) = (0.5957, 0.7015, 0.7865)$. For the Wratten #12 band and MOA-*Red* band data, we approximately take $u \sim (u_I + u_R)/2$.

The best-fit parameters with their uncertainties are listed in Table 5. The modeling indicates that B is the best solution, and (A, C, D, E, F) are disfavored by $\Delta\chi^2 = (0.1, 3.4, 2.5, 83.4, 83.3)$. The model light curves together with the data are shown in Fig. 4. We rule out the binary star interpretations (Solutions E and F) because they failed to describe the light curve with relatively large $\Delta\chi^2$. We note the similarity in the degeneracy between solution pairs (A, B) and (C, D) with that in Section 4.2.1. This will be further discussed in Section 7.2. The caustic structure of each solution is shown in Figure 9.

We further investigate the microlens parallax effect. The parallax fitting improves the solution by $\Delta\chi^2 \sim 17$ for A and B, $\Delta\chi^2 \sim 19$ for C and D, $\Delta\chi^2 \sim 22$ for $u_0 < 0$ of E and F, and $\Delta\chi^2 \sim 32$ for $u_0 > 0$ of E and F. All the solutions give a $2\sigma$ lower-limit on the parallax of at least $\pi_{\rm E} > 1.3$. However, for a relatively short $t_{\rm E} \sim 23$ d event, the detection of microlens parallax is not common. After a further investigation, we finally ruled out the microlens parallax detection for two reasons. First, the parallax signals are only from the two KMTC datasets, and the signal trends versus time do not match up with the other sites. Second, and more importantly, the baseline data dominate the parallax signal, whereas the peak data provides no signal. These two factors suggest that the parallax signal is caused by unknown systematic errors, and we therefore rule out the parallax detections.

#### 4.3.2 Binary-source (1L2S) modelling

We also search for 1L2S solutions for KMT-2021-BLG-1689 using MCMC. The parameters of the best fit model is shown in Table 6. Although Figure 4 indicates that the 1L2S model describes the light curve reasonably well, it is disfavored by the following reasons.

(1). Despite 1L2S model having three additional parameters, the $\chi^2$ is 25.9 larger than the best 2L1S model.

(2). We follow a similar procedure as Han et al. (2022) used for KMT-2021-BLG-0240 (see their Section 3.4). From Section 5 we measure the angular radius of the first source, $\theta_{*,1} \sim 0.54~\mu{\rm as}$. Thus the projected physical separation of the two sources is

$$r_{s,\perp} \equiv r_s \sin i = D_S \frac{\theta_{*,1}}{\rho_1} \Delta u, \quad (15)$$

$$\Delta u = \sqrt{\left(\frac{t_{0,1} - t_{0,2}}{t_E}\right)^2 + (u_{0,1} - u_{0,2})^2} = 0.00805 \pm 0.00053. \quad (16)$$

Where $D_S$ is the distance to the sources from Earth, and $i$ is the angle between the line-of-sight and the orbital plane. Assuming the mass of two sources are $M_{s,1} = 0.5 M_\odot$ and $M_{s,2} = 0.1 M_\odot$ and the source distance $D_S = 8.3$ kpc, we estimate the orbital period $P$ of the two sources by sampling over the angle $i$. We find $P = 1.1^{+5.8}_{-0.7}$ days $\lesssim 0.1 t_E$. Moreover, the position change in the unit of $\theta_E$ of the primary source during $P/2$ is $2(M_{s,1}/M_{s,2}) \Delta u \sim 0.080 \gg u_0$. With this relatively short period and large positional change, the light curve would show violent changes by the orbital motion of the two sources as the microlens "xallarap" effect. However, no such signals were observed on the light curve. (See, for example, Figure 3 of Han & Gould (1997) for an illustration of the xallarap effect in a light curve.)

We therefore rule out the 1L2S interpretation.

## 5 SOURCE PROPERTIES

The purpose of this section is to measure the color of the source star. The color, on the one hand, allows us to estimate the $T_{\rm eff}$ and the limb-darkening coefficients in Section 4, and on the other hand, can be used to measure the angular radius of the source star, $\theta_*$. With the source radius, we can measure

$$\theta_E = \frac{\theta_*}{\rho}, \quad \mu_{\rm rel} = \frac{\theta_E}{t_E}, \quad (17)$$

which are directly related to the physical parameters of the lens.

For the first step, we place the source on CMD using the KMTNet data. Then we measure the offset of the source relative to the centroid of the red clump giants (Yoo et al. 2004),

$$\Delta[(V - I), I] = [(V - I), I]_s - [(V - I), I]_{\rm RC}. \quad (18)$$

By comparing the instrumental $[(V - I), I]_{\rm RC}$ with the intrinsic centroid of the red giant clump $[(V - I), I]_{\rm RC,0}$ from Bensby et al. (2013) and Nataf et al. (2013), we can find the intrinsic, de-reddened color and magnitude of the source

$$[(V - I), I]_{s,0} = [(V - I), I]_{\rm RC,0} + \Delta[(V - I), I]. \quad (19)$$

Based on the de-reddened color and magnitude, we estimate the source angular radius $\theta_*$ from Adams et al. (2018). We also estimate the effective temperature $T_{\rm eff}$ of the source (Houdashelt et al. 2000) to determine the limb-darkening coefficients used in Section 4.

For both events, we construct CMDs from stars within a $2' \times 2'$ square centered on the source position using KMTC01 data. The CMDs are shown in Figure 10. The source color is determined from the regression of the $V$-band and $I$-band source fluxes during the event. The de-reddened source color together with the derived parameters are listed in Table 7.

## 6 LENS PROPERTIES

Our objective in this section is to estimate the physical parameters of the lens. If both $\theta_E$ and $\pi_E$ are measured in the light curve, the lens mass can be directly derived by

$$M_L = \frac{\theta_E}{\kappa \pi_E}, \quad \kappa = \frac{4G}{c^2 \rm AU} \simeq 8.144 \, {\rm mas}/M_\odot, \quad (20)$$

where $G$ is the gravitational constant and $c$ is the speed of light. However, for KMT-2021-BLG-0171 we only measure one-dimensional $\pi_E$, and for KMT-2021-BLG-1689 we do not measure the parallax.

Therefore, we estimate the lens physical parameters from a Bayesian analysis using the Galactic model as priors. We adopt the Galactic "Model C" described in Yang et al. (2021). We generate a large number of simulated microlensing events based on the Galactic model, that is, generating source and lens distance from the line-of-sight stellar density profiles, generating lens mass from the mass function, and generating source and lens motions from the stellar velocity distribution. The prior is based on the assumption that the probability of a star to host a planet is independent of its mass and Galactic environment. For each simulated event, $i$, we weight it by

$$w_i = \Gamma_i \times \mathcal{L}_i(t_E) \mathcal{L}_i(\theta_E) \mathcal{L}_i(\pi_{E,i}), \quad (21)$$

where $\Gamma_i \propto \theta_{E,i} \mu_{{\rm rel},i}$ is the event rate. $\mathcal{L}(t_E)$, $\mathcal{L}(\theta_E)$, and $\mathcal{L}(\pi_E)$ are the likelihood function measured from a specific solution in Section 4.

### 6.1 KMT-2021-BLG-0171

We generate $2 \times 10^9$ simulated events according to the Galactic model and weight them by Eq. 21, where the $t_E$, $\theta_E$ and $\pi_E$ constraints are derived from the fits. Because the two components of $\pi_E$ are not independent, the full covariances are used. The angle of the minor axis of the error ellipse (north through east) is $\psi \sim 95°$ for all solutions. However, many of the simulated events have small $D_L$ and large $M_L$, which is in conflict with the observed blend flux. We measure the baseline blend light in the CFHT images to be $I_b = 19.33 \pm 0.07$ (Zang et al. 2018). The lens flux should not be brighter than the blend light, thus we set an upper limit of the blend flux to be $I_{b,{\rm limit}} = 19.12$ ($3\sigma$). Therefore we reject simulated events that have too bright lens hosts. For main sequence lens stars, the $I$-band absolute magnitude $M_I$ is a function of mass (Wang et al. 2018),

$$M_I = 4.4 - 8.5 \log \frac{M}{M_\odot}. \quad (22)$$

The rejection threshold is

$$M_I + 5 \log \frac{D_L}{10 \, {\rm pc}} > I_{b,{\rm limit}} - A_I(D_L), \quad (23)$$

Where $A_I(D_L)$ is the $I$-band extinction for a lens star in given distance $D_L$,

$$A_I(D_L) = \int_0^{D_L} a_I \times n_d(D) \, dD, \quad (24)$$

where $n_d(D)$ is the dust number density at given distance $D$, and $a_I$ is a constant which describes the extinction caused by per kpc$^{-3}$ dust. We adopt the exponential Galactic dust distribution as follows

$$n_d(D) \propto e^{-\frac{|z(D)|}{z_d} - \frac{R(D)}{R_d}}, \quad \text{where} \quad (25)$$

$$z(D) = z_\odot + D \sin b \approx z_\odot + Db, \quad (26)$$

$$R(D) = \sqrt{(R_\odot - D \cos b \cos l)^2 + (D \cos b \sin l)^2} \quad (27)$$
$$\approx |R_\odot - D|.$$

Here $(R, z)$ are the axis of Galactic cylindrical coordinates, and $(R_\odot, z_\odot) = (8.3, 0.023)$ kpc is the location of the Sun. We adopt the dust scale lengths from Li et al. (2018), where $(R_d, z_d) = (3.2, 0.1)$. We determine the extinction constant $a_I = 4.13$ by applying $A_I(D_S) = I_{\rm RC} - I_{\rm RC,0}$ measured in Section 5 and assuming $D_S = 8.3$ kpc. The result is not sensitive to the $D_S$ value for bulge sources because most of the extinction occurs near the Galactic disk plane.

We weight the remaining events by Eq. 21. The final results of the physical parameters are shown in Table 8. We combine the results from all solutions by weighting each solution by $e^{-\Delta\chi^2/2}$. The combined distribution of the host and the planet parameters are shown in Figure 12. The blended light limit is plotted with the magenta dashed line.

The results indicate that the lens star is likely to be a $\sim 0.8 M_\odot$ K-type star. For the A & B solutions, the planet has a mass $\sim 12 M_\oplus$ and is orbiting at a projected separation of $\sim 2.9$ AU or $\sim 4.5$ AU, respectively. For the somewhat disfavored C & D solutions, the planet has a mass $\sim 6 M_\oplus$ and is orbiting at a projected separation of $\sim 3.7$ AU. The planetary system is more likely to be located in the Galactic disk at $D_L \sim 4.4 - 5$ kpc from our solar system. In addition, from Figure 12, we note that the host has $\sim 12\%$ chance of being a white dwarf (based on the assumption that white dwarfs have the same probability as main-sequence stars to host such a planet). If the host is a main-sequence star, the Bayesian results predict a brightness $I = 19.9^{+0.9}_{-0.6}$ and a $3\sigma$ limit $I < 22.9$.

We also checked the astrometric alignment between the source and the baseline object from KMTC images and CFHT images (Zang et al. 2018). The astrometric offset between the source and the baseline object is

$$\Delta\theta(N, E) = (8 \pm 6, 3 \pm 5) \text{mas}. \quad (28)$$

Therefore, the baseline object is consistent with the position of the event at the $\sim 1\sigma$ level. Thus, the lens could account for most or all of the blend light.

The alignment can be immediately checked (e.g., 2022 season) by the *Hubble Space Telescope (HST)* or by ground-based large telescopes equipped with adaptive optics instruments (e.g., Keck, Subaru). However, if further observation finds the blend light is from other, unrelated, stars, i.e., the lens is much fainter than expected, the white dwarf lens interpretation would be preferred.

### 6.2 KMT-2021-BLG-1689

We generate $5\times10^7$ simulated events according to the Galactic model and weight them by Eq. 21. The likelihood function $\mathcal{L}_i(\pi_{E,i})$ is set to be a constant because we do not measure the parallax for this event. In addition, because the blended light does not provide extra limits for this event, we keep all the simulated events. The results of the physical parameters from the Bayesian analysis are listed in Table 9. We also combine all the solutions by their $\chi^2$, the combined distribution is shown in the upper panels of Figure 13. Solution C and D become negligible after the weighting because they are disfavored by both $\chi^2$ and the Galactic model. We separately display the result distribution for Solution C and D in the bottom panels of Figure 13. Solution (A,B) and (C,D) predict greatly different $\theta_E$ and thus greatly different $\mu_{\rm rel}$, which can be tested by future high-resolution imaging follow-up observations.

If Solution A or B is correct, the results imply that the lens is likely to be a $\sim 0.6 M_\odot$ M dwarf located in the Galactic bulge ($\sim 7.2$ kpc), and the planet, with mass $\sim 46 M_\oplus$ is orbiting it at a distance of $2-3$ AU. For Solutions C and D, the lens is likely to be a $\sim 0.7 M_\odot$ star in the Galactic disk ($\sim 5.0$ kpc), and a $\sim 39 M_\oplus$ planet is orbiting it at a distance of $\sim 3.3$ AU. In both interpretations, the planet mass, $\sim 30 - 40 M_\oplus$, is located in the runaway accretion "desert" (Ida & Lin 2004).

Moreover, the white dwarf interpretation ($\sim 8\%$ probability) can be tested by future high resolution imaging followup. The Bayesian results predict a brightness of a main-sequence host to be $I = (22.6^{+2.0}_{-1.7}, 20.8^{+1.7}_{-1.2})$ for (A/B, C/D), respectively.

## 7 DISCUSSION

### 7.1 Role of the HighMagFinder and followup

We have shown that HighMagFinder is effective at identifying and alerting high-magnification microlensing events in time for followup observations. Initial trials of the HighMagFinder algorithm show that KMT-2021-BLG-0171 would have been alerted at least as soon as it was identified by eye as a high-magnification event. Later, KMT-2021-BLG-1689 was alerted during regular operation of the HighMagFinder as a high-magnification event, leading to crucial followup observations characterizing the planetary perturbation in this event.

To further quantify the role of followup observations, we repeat the analysis using only survey data (from KMTNet and MOA) and compare to the results when followup data are included.

For KMT-2021-BLG-1689, we find that with survey data only, the planetary signal cannot be well characterized. The goodness of the 1L2S solution is comparable to the 2L1S solution ($\Delta\chi^2 \sim 6.8$). Moreover, there is no other evidence that strongly disfavors the 1L2S interpretation (The uncertainty of the xarallap interpretation becomes larger). All the 2L1S solutions in Section 4.3 can still fit the light curve, and the parameters are consistent at $2\sigma$ with those in Section 4.3. However, the solutions are more degenerate. For instance, the binary star E and F solutions are only disfavored by $\Delta\chi^2 \sim 22$. To summarize, the Auckland Observatory followup data of KMT-2021-BLG-1689 helped us resolve the degeneracies between 2L1S and 1L2S solutions and between 2L1S planetary and stellar binary interpretations. Thus the followup data are essential for the discovery of this planet.

For KMT-2021-BLG-0171, the planet can still be well characterized without followup data. The 1L2S and 2L1S models can describe the light curve almost equally well ($\Delta\chi^2 \sim 0$), but the 1L2S interpretation can still be ruled out by following the approach in Section 4.2.1. However, the uncertainty of the 2L1S parameters are larger. For example, we measure the mass ratio in Solution $A_-$ to be $q = (5.45 \pm 1.88) \times 10^{-5}$, i.e., with an uncertainty that is about a factor of two larger than the one shown in Table 3.

In addition to the planet that was actually detected in this event, we show that the followup data generally make the light curve more sensitive to even smaller planets. The planetary sensitivity of a microlensing event is defined as the probability to detect the planetary signals if the lens hosts a given $(\log s, \log q)$ planet. We follow the methods described in Suzuki et al. (2016) to calculate the sensitivity for KMT-2021-BLG-0171 with and without the followup data. We set the detection threshold to be $\Delta\chi^2_{\rm threshold} = 200$, and sample over $(-0.3 \leqslant \log s \leqslant 0.3, -6.0 \leqslant \log q \leqslant -3.0, 0° \leqslant \alpha < 360°)$ with $(31, 31, 360)$ values. The results are shown in Figure 14. A binned (over $\log q$) sensitivity is shown in Table 10. The follow-up data enlarges the sensitivity significantly. The sensitivity as a function of $\log q$ is extended by about 0.4 dex toward smaller $q$, which is essential for searching smaller planets.

### 7.2 Degeneracies

The well-understood degeneracies of 2L1S microlensing light curves are mostly "intrinsic" degeneracies. The intrinsic degeneracies are caused by the symmetry of the lens equation and can result in in-trinsically similar magnification maps and light curves. "Intrinsic" means the similarity is almost independent of the data sampling.

For high-magnification microlensing events, the anomalies are mainly caused by central or resonant caustics. Thus, the degeneracy in central caustic morphologies can cause the degeneracy in the light curves. The well-know "close-wide" degeneracy, which approximately takes $s \leftrightarrow s^{-1}$ (Griest & Safizadeh 1998; Dominik 1999; An 2005), is derived in this way. For the two events reported in this paper, the degeneracy between Solutions A and B are well described by the "close-wide" degeneracy.

When the planetary caustic creates the anomaly on the light curves, another degeneracy called "inner-outer" degeneracy emerges (Gaudi & Gould 1997). When the planetary caustic is small, the source passing by different sides of the caustic can create similar light curves.

Recently, it was realized that the "close-wide" and "inner-outer" degeneracies can be unified as a more general degeneracy and can be extended to the resonant region (Zhang et al. 2021; Ryu et al. 2022). The degeneracy is related to the trajectory by $u_{\rm anom}$,

$$u_{\rm anom} = \sqrt{u_0^2 + \left(\frac{t_{\rm anom} - t_0}{t_{\rm E}}\right)^2} = \left|\frac{u_0}{\sin \alpha}\right| \quad (29)$$

where $t_{\rm anom}$ is the time of the anomaly signal or the time when the source crosses the line connecting the two lenses. We have $u_{\rm anom} \sim 0.0067$ for both events. For two degenerate solutions with similar $q$ but different separations $s_1$ and $s_2$, Ryu et al. (2022) suggest that

$$s_{\pm}^{\dagger} \equiv \frac{\sqrt{u_{\rm anom}^2 + 4} \pm u_{\rm anom}}{2} = \sqrt{s_1 s_2}. \quad (30)$$

For anomalous bumps, we take the "+" sign. For the central caustic cases, $s_{+}^{\dagger} \simeq 1$ and the formula becomes the "close-wide" degeneracy, $s_1 s_2 \sim 1$. We find $s_{+}^{\dagger} \sim 1.0033$ for both events. For KMT-2021-BLG-(0171,1689), we find $\sqrt{s_A s_B} = (1.0009, 1.0027)$ and $\sqrt{s_C s_D} = (1.0032, 1.0033)$, which are consistent with Eq.30.

Slightly differently, Zhang et al. (2021) suggests that approximately,

$$u_{\rm anom} \equiv x_{\rm null} = \frac{1}{2}(s_1 - \frac{1}{s_1} + s_2 - \frac{1}{s_2}). \quad (31)$$

Similarly, when $u_{\rm anom} \sim 0 \ll 1$, we have $1/s_1 \gg s_1$ and $s_2 \gg 1/s_2$ for the "close" and "wide" solutions, respectively, the equation becomes the "close-wide" degeneracy, $1/s_1 - s_2 \sim 0$ or $s_1 s_2 \sim 1$. We find $x_{\rm null,AB} = (0.0018, 0.0055)$ and $x_{\rm null,CD} = (0.0064, 0.0066)$ for KMT-2021-BLG-(0171,1689). Overall, both Eq.30 and Eq.31 can reproduce the "intrinsic" degeneracy between Solutions A(C) and B(D).

In addition to the "intrinsic" degeneracy, some other degeneracies are accidentally caused by the data sampling. The degeneracy between (A,B) and (C,D) for KMT-2021-BLG-1689 belongs to this type of degeneracy. The solutions (A,B) and (C,D) predict different source radii, $\rho$, and mass ratios, $q$: the anomaly is explained by a large source crossing the central caustic or a smaller source crossing a resonant/near-resonant caustic. As a result, the light curve of the anomaly signal could have either a single-peak or double-peak feature (see Figure 4). Similar to Yee et al. (2021), better sampling or more accurate data could help to resolve this degeneracy. In addition, as shown in Tables 5 and 7, the two sets of solutions predict greatly different $\rho$ and consequently different $\theta_{\rm E}$, which can be distinguished by future high-resolution follow-up observations. A similar $\rho - q$ degeneracy is also found in MOA-2011-BLG-262 (Bennett et al. 2014), KMT-2021-BLG-1391 and KMT-2021-BLG-1253 (Ryu et al. 2022).

However, for KMT-2021-BLG-0171, it is hard to tell whether the degeneracy between (A,B) and (C,D) is "intrinsic" or "accidental". It would seem that the mechanism for this degeneracy is the same as the above $\rho - q$ degeneracy, but in this case, the degenerate solutions predict almost identical $\rho$, and the light curves of the anomaly in all solutions are single-peaked. This means the solutions can be distinguished by neither better sampling nor future follow-up observations. We searched the literature and found a similar case, OGLE-2011-BLG-0251 (Kains et al. 2013), but the resonant solution was excluded.

Despite the fact that the (A,B)-(C,D) degeneracy in the two events appears to be somehow different and is not well-understood, we can draw some general inferences from their similarites. First, combined with KMT-2021-BLG-1391 and KMT-2021-BLG-1253 (Ryu et al. 2022) mentioned above, we find four events suffer this "central caustic" - "resonant caustic" degeneracy in 2021. This indicates that similar degenerate solutions might have been missed in previous events and suggests that we should explore the parameter space more carefully in future events. Second, the magnification map as a function of $s$ varies rapidly in the resonant or near-resonant region. In general, to prevent missing possible solutions, we should pay more attention to this region when searching for solutions (e.g., operating a grid search).

Finally, the resonant or near-resonant region is also important when calculating the sensitivity. We show a zoom-in of the sensitivity plot Figure 14 with denser $\log s$ grids in Figure 15. The refined calculation suggests that the sensitivity for Solutions C and D (the two crosses within the resonant region) is nearly 100%. However, if we estimate from the interpolation of Figure 14, the sensitivity would be $\sim 70\%$. Underestimation of sensitivity can lead to overestimation of the occurrence rate of such planets. As a result, statistical studies should also pay more attention to the resonant regions.

### 7.3 Phase-space Consideration of the 2L1S Solutions

From Figures 5 and 8 and Tables 2 and 5, we notice that the coverage of all solutions in ($\log s$, $\log q$, $\alpha$) space are different. As a prior, the intrinsic distributions of planets in $\log s$, $\log q$, and $\alpha$ should be uniform or nearly uniform. So the solution that covers larger phase space would be more likely to be true.

More quantitatively, we estimate the phase-space factor of each solution from the MCMC chains. First, we calculate the covariance matrix of these parameters from the chain,

$$C_{ij} = {\rm cov}(a_i, a_j), \quad a_i, a_j = (\log s, \ \log q, \ \alpha). \quad (32)$$

By assuming that the the distribution is approximately multi-Gaussian, the phase-space factor of a solution is then

$$p = \sqrt{\det(C)}. \quad (33)$$

For KMT-2021-BLG-0171, we find $p_A : p_B : p_C : p_D \approx 60 : 63 : 1 : 1$, which is equivalent to $\Delta\chi^2 \approx (0.1, 0.0, 8.3, 8.3)$. Thus the resonant solutions C and D are strongly disfavored under this consideration. We can also include the mass-ratio function factor as a prior: $dN_{\rm pl}/d\log q \propto q^{-\gamma}$. For example, if we choose $\gamma = 0.6$ (Gould et al. 2010), then the overall phase-space factor is $p_A : p_B : p_C : p_D \approx 40 : 43 : 1 : 1$.

As for KMT-2021-BLG-1689, we obtain $p_A : p_B : p_C : p_D \approx 15.0 : 14.3 : 1.1 : 1$ and $p_A : p_B : p_C : p_D \approx 12.7 : 12.1 : 1.1 : 1$ with and without the mass-ratio function prior, respectively.

The phase-space factor is independent from $\Delta\chi^2$. In both events, the resonant solutions are unlikely to be true because they only occupy small regions in the phase space. We argue that the phase-space factors should be included in future statistical studies to weight degenerate solutions (together with $\Delta\chi^2$). Therefore, for the two events reported in this paper, although the resonant solutions describe the light-curves reasonably well, they may not be statistically important.

## ACKNOWLEDGEMENTS

H.Y., W.Zang, J.Z., S.M. and W.Zhu acknowledge support by the National Natural Science Foundation of China (Grant No. 12133005). This research has made use of the KMTNet system operated by the Korea Astronomy and Space Science Institute (KASI) and the data were obtained at three host sites of CTIO in Chile, SAAO in South Africa, and SSO in Australia. This research uses data obtained through the Telescope Access Program (TAP), which has been funded by the TAP member institutes. The MOA project is supported by JSPS KAKENHI grant Nos. JSPS24253004, JSPS26247023, JSPS23340064, JSPS15H00781, JP16H06287, JP17H02871, and JP19KK0082. This research is supported by Tsinghua University Initiative Scientific Research Program (Program ID 2019Z07L02017). J.C.Y. acknowledges support from N.S.F Grant No. AST-2108414. Work by C.H. was supported by the grants of National Research Foundation of Korea (2019R1A2C2085965 and 2020R1A4A2002885). W.Zhu acknowledges the science research grants from the China Manned Space Project with No. CMS-CSST-2021-A11. The authors acknowledge the Tsinghua Astrophysics High-Performance Computing platform at Tsinghua University for providing computational and data storage resources that have contributed to the research results reported within this paper.

## DATA AVAILABILITY

Data used in the light curve analysis will be provided along with publication.

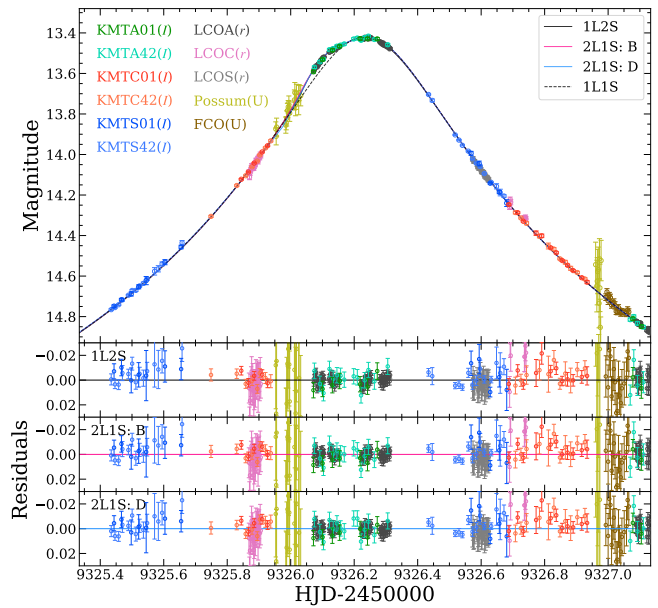

**Figure 3.** Light curve data of KMT-2021-BLG-0171 around the peak together with the best-fit models: binary-lens "B", binary-lens "D" and binary-source (1L2S). The residuals for each model are shown in separate panels. For clarity, we only plot Models B and D for binary-lens models, because Models A (C) and B (D) are not visibly different. The model light curve and data have been aligned to the KMTC $I$-band magnitude. The names and filters for each dataset are labeled on the panel, where $U$ means unfiltered.

This paper has been typeset from a TEX/LATEX file prepared by the author.

**Table 2.** Static 2L1S models for KMT-2021-BLG-0171

| Solution | $t_0$ (HJD′) | $u_0$ | $t_E$ (d) | $\rho$ ($10^{-3}$) | $\alpha$ (rad) | $s$ | $q$ ($10^{-4}$) | $I_s$ | $\chi^2$/dof |
|---|---|---|---|---|---|---|---|---|---|
| A | 9326.2338 | 0.00564 | 41.57 | 0.150 | 4.147 | 0.813 | 4.28 | 19.05 | 3728.2/3728 |
|   | 0.0003 | 0.00005 | 0.32 | 0.015 | 0.012 | 0.032 | 0.80 | 0.01 | |
| B | 9326.2338 | 0.00564 | 41.56 | 0.151 | 4.149 | 1.232 | 4.17 | 19.05 | 3728.4/3728 |
|   | 0.0003 | 0.00005 | 0.32 | 0.015 | 0.012 | 0.051 | 0.82 | 0.01 | |
| C | 9326.2338 | 0.00565 | 41.57 | 0.162 | 4.173 | 0.9905 | 2.19 | 19.05 | 3731.9/3728 |
|   | 0.0003 | 0.00005 | 0.32 | 0.007 | 0.007 | 0.0009 | 0.14 | 0.01 | |
| D | 9326.2338 | 0.00565 | 41.55 | 0.162 | 4.171 | 1.0161 | 2.22 | 19.05 | 3733.9/3728 |
|   | 0.0003 | 0.00005 | 0.31 | 0.007 | 0.007 | 0.0009 | 0.15 | 0.01 | |

NOTE. HJD′ = HJD − 2450000.

**Table 3.** Parallax 2L1S models for KMT-2021-BLG-0171

| Solution | $t_0$ (HJD′) | $u_0$ | $t_E$ (d) | $\rho$ ($10^{-3}$) | $\alpha$ (rad) | $s$ | $q$ ($10^{-4}$) | $\pi_{E,N}$ | $\pi_{E,E}$ | $I_s$ | $\chi^2$/dof |
|---|---|---|---|---|---|---|---|---|---|---|---|
| $A_+$ | 9326.2339 | 0.00568 | 41.36 | 1.48 | 4.146 | 0.801 | 0.464 | −0.093 | −0.043 | 19.05 | 3719.7/3726 |
|   | 0.0003 | 0.00005 | 0.32 | 0.17 | 0.011 | 0.035 | 0.097 | 0.175 | 0.020 | 0.01 | |
| $A_-$ | 9326.2338 | −0.00560 | 41.43 | 1.49 | 2.139 | 0.798 | 0.479 | −0.332 | −0.063 | 19.05 | 3718.0/3726 |
|   | 0.0003 | 0.00007 | 0.34 | 0.17 | 0.011 | 0.034 | 0.096 | 0.243 | 0.024 | 0.01 | |
| $B_+$ | 9326.2338 | 0.00568 | 41.37 | 1.50 | 4.146 | 1.247 | 0.450 | −0.070 | −0.041 | 19.04 | 3720.0/3726 |
|   | 0.0003 | 0.00006 | 0.34 | 0.16 | 0.011 | 0.056 | 0.094 | 0.188 | 0.020 | 0.01 | |
| $B_-$ | 9326.2338 | −0.00561 | 41.38 | 1.46 | 2.135 | 1.263 | 0.474 | −0.298 | −0.060 | 19.05 | 3718.5/3726 |
|   | 0.0003 | 0.00007 | 0.33 | 0.16 | 0.010 | 0.054 | 0.095 | 0.256 | 0.025 | 0.01 | |
| $C_+$ | 9326.2339 | 0.00571 | 41.32 | 1.62 | 4.174 | 0.9906 | 0.220 | −0.157 | −0.044 | 19.05 | 3724.8/3726 |
|   | 0.0003 | 0.00005 | 0.34 | 0.06 | 0.007 | 0.0009 | 0.014 | 0.176 | 0.018 | 0.01 | |
| $C_-$ | 9326.2337 | −0.00566 | 41.37 | 1.63 | 2.109 | 0.9905 | 0.222 | −0.154 | −0.045 | 19.05 | 3724.4/3726 |
|   | 0.0003 | 0.00007 | 0.33 | 0.06 | 0.007 | 0.0008 | 0.014 | 0.250 | 0.024 | 0.01 | |
| $D_+$ | 9326.2339 | 0.00571 | 41.34 | 1.61 | 4.171 | 1.0160 | 0.221 | −0.185 | −0.046 | 19.04 | 3726.6/3726 |
|   | 0.0004 | 0.00006 | 0.34 | 0.06 | 0.006 | 0.0010 | 0.015 | 0.175 | 0.020 | 0.01 | |
| $D_-$ | 9326.2337 | −0.00565 | 41.39 | 1.63 | 2.112 | 1.0162 | 0.224 | −0.137 | −0.043 | 19.04 | 3726.8/3726 |
|   | 0.0003 | 0.00007 | 0.32 | 0.06 | 0.007 | 0.0010 | 0.014 | 0.250 | 0.023 | 0.01 | |

NOTE. HJD′ = HJD − 2450000.

**Table 4.** 1L2S models for KMT-2021-BLG-0171

| Dataset | $t_{0,1}$ | $t_{0,2}$ | $u_{0,1}$ | $u_{0,2}$ | $t_E$ (d) | $10^3 \rho_1$ | $10^3 \rho_2$ | $q_{f,I}$ | $q_{f,r}$ | $q_{f,U}$ | $q_{f,V}$ | $I_{s,1}$ | $\chi^2$/dof |
|---|---|---|---|---|---|---|---|---|---|---|---|---|---|
| Fiducial | 9326.2375 | 9326.0941 | 0.0057 | 0.0000 | 41.72 | <4.6 | 1.47 | 0.0065 | 0.0063 | <0.057 | − | 19.06 | 3739.8 |
|   | 0.0004 | 0.0029 | 0.0001 | 0.0005 | 0.31 | − | 0.11 | 0.0009 | 0.0009 | − | − | 0.01 | / 3725 |
| Fiducial + KMTNet $V$ | 9326.2377 | 9326.0934 | 0.0057 | −0.0001 | 41.78 | <4.4 | 1.51 | 0.0068 | 0.0065 | <0.065 | 0.0067 | 19.07 | 4169.8 |
|   | 0.0005 | 0.0027 | 0.0001 | 0.0006 | 0.32 | − | 0.13 | 0.0014 | 0.0013 | − | 0.0019 | 0.01 | / 4144 |

NOTE. $t_{0,1}$ and $t_{0,2}$ are in HJD′, where HJD′ = HJD − 2450000.

**Table 5.** Static 2L1S models for KMT-2021-BLG-1689

| Solution | $t_0$ | $u_0$ | $t_E$ (d) | $\rho$ ($10^{-3}$) | $\alpha$ (rad) | $s$ | $q$ ($10^{-4}$) | $I_s$ | $\chi^2$/dof |
|---|---|---|---|---|---|---|---|---|---|
| A | 9409.2510 | 0.00600 | 22.56 | 1.44 | 4.230 | 0.870 | 2.10 | 21.60 | 9060.4/9057 |
|   | 0.0011 | 0.00028 | 0.84 | 0.08 | 0.010 | 0.025 | 0.39 | 0.04 | |
| B | 9409.2509 | 0.00601 | 22.51 | 1.44 | 4.229 | 1.157 | 2.09 | 21.60 | 9060.3/9057 |
|   | 0.0011 | 0.00026 | 0.79 | 0.08 | 0.010 | 0.032 | 0.37 | 0.04 | |
| C | 9409.2509 | 0.00590 | 22.61 | 0.70 | 4.226 | 0.944 | 1.62 | 21.61 | 9063.7/9057 |
|   | 0.0012 | 0.00027 | 0.85 | 0.08 | 0.010 | 0.004 | 0.17 | 0.04 | |
| D | 9409.2510 | 0.00587 | 22.78 | 0.68 | 4.228 | 1.067 | 1.62 | 21.60 | 9062.8/9057 |
|   | 0.0011 | 0.00027 | 0.81 | 0.08 | 0.009 | 0.005 | 0.18 | 0.04 | |
| E | 9409.2403 | 0.00663 | 22.92 | < 1.2 | 5.950 | 0.092 | 5079 | 21.62 | 9143.7/9057 |
|   | 0.0012 | 0.00032 | 0.88 | − | 0.017 | 0.006 | 2232 | 0.05 | |
| F | 9409.2394 | 0.00327 | 46.14 | < 0.8 | 2.807 | 19.97 | 3186 | 21.63 | 9143.6/9057 |
|   | 0.0009 | 0.00060 | 8.48 | − | 0.009 | 1.25 | 1979 | 0.05 | |

NOTE. HJD′ = HJD − 2450000. The values of the parameters $t_0$ and $u_0$ are with respect to the different origins for different solutions. In (A, B, C), the origin is the center of mass $x_{\rm mass}$. In (B, D), the origin is taken to be the magnification center of the primary lens, where $x_{\rm mag,1} = x_{\rm mass} - (s - s^{-1})q/(1+q)$. In Solution F, the origin is set to the magnification center of the secondary lens, where $x_{\rm mag,2} = x_{\rm mass} + (s - s^{-1})/(1+q)$.

**Table 6.** 1L2S model for KMT-2021-BLG-1689

| Dataset | $t_{0,1}$ | $t_{0,2}$ | $u_{0,1}$ | $u_{0,2}$ | $t_E$ (d) | $\rho_1$ ($10^{-3}$) | $\rho_2$ ($10^{-3}$) | $q_{f,I}$ | $q_{f,U}$ | $q_{f,R}$ | $I_{s,1}$ | $\chi^2$/dof |
|---|---|---|---|---|---|---|---|---|---|---|---|---|
| Fiducial | 9409.2615 | 9409.1833 | 0.0072 | 0.0000 | 22.12 | < 8.0 | 1.47 | 0.0527 | 0.0674 | 0.0418 | 21.62 | 9086.2 |
|  | 0.0023 | 0.0009 | 0.0004 | 0.0003 | 0.79 | – | 0.08 | 0.0185 | 0.0101 | 0.0039 | 0.05 | / 9054 |

NOTE. $t_{0,1}$ and $t_{0,2}$ are in HJD$'$, where HJD$'$ = HJD − 2450000.

**Table 7.** Source properties and drived $\theta_E$, $\mu_{\rm rel}$ for the two events

| Event | KMT-2021-BLG-0171 | KMT-2021-BLG-1689 |
|---|---|---|
| $[(V-I), I]_{\rm RC}$ | [2.307±0.013, 16.16±0.08] | [2.46±0.04, 16.24±0.14] |
| $[(V-I), I]_{\rm s}$ | [2.119±0.003, 19.05±0.01] | [2.58±0.04, 21.59±0.04] |
| $[(V-I), I]_{\rm RC,0}$ | [1.06, 14.430] | [1.06, 14.426] |
| $[(V-I), I]_{\rm s,0}$ | [0.872±0.013, 17.32±0.08] | [1.18±0.05, 19.77±0.15] |
| $\theta_*$ ($\mu$as) | 1.28±0.08 | 0.54±0.06 |
| $T_{\rm eff}$ (K) | ~5200 | ~4570 |
| $\theta_E$ (mas) | 0.86±0.11  for $A_\pm$, $B_\pm$ | 0.37±0.04  for $A$, $B$ |
|  | 0.79±0.06  for $A_\pm$, $B_\pm$ | 0.77±0.12  for $A$, $B$ |
| $\mu_{\rm rel}$ (mas/yr) | 7.6±1.1  for $A_\pm$, $B_\pm$ | 6.1±0.8  for $A$, $B$ |
|  | 7.0±0.6  for $A_\pm$, $B_\pm$ | 12.6±2.0  for $A$, $B$ |

**Table 8.** Physical parameters from Bayesian analysis for KMT-2021-BLG-0171

| Solution | Physical Properties | | | | | Relative Weights | | |
|---|---|---|---|---|---|---|---|---|
|  | $M_{\rm host}$ ($M_\odot$) | $M_{\rm planet}$ ($M_\oplus$) | $D_L$ (kpc) | $r_\perp$ (AU) | $\mu_{\rm rel}$ (mas/yr) | Gal. Mod. | $\chi^2$ | Total |
| $A_+$ | $0.80^{+0.28}_{-0.24}$ | $12.1^{+5.2}_{-4.2}$ | $4.7^{+1.7}_{-1.1}$ | $2.9^{+0.6}_{-0.5}$ | $6.7^{+0.9}_{-0.8}$ | 0.728 | 0.421 | 0.981 |
| $A_-$ | $0.76^{+0.31}_{-0.30}$ | $11.8^{+5.7}_{-4.9}$ | $4.5^{+1.8}_{-1.3}$ | $2.8^{+0.7}_{-0.7}$ | $6.7^{+1.0}_{-0.8}$ | 0.263 | 1.000 | 0.840 |
| $B_+$ | $0.78^{+0.27}_{-0.24}$ | $11.5^{+4.9}_{-4.0}$ | $4.6^{+1.5}_{-1.0}$ | $4.5^{+0.9}_{-0.8}$ | $6.8^{+0.9}_{-0.8}$ | 0.844 | 0.370 | 1.000 |
| $B_-$ | $0.74^{+0.29}_{-0.27}$ | $11.4^{+5.4}_{-4.5}$ | $4.4^{+1.6}_{-1.2}$ | $4.4^{+1.0}_{-1.1}$ | $6.9^{+0.9}_{-0.8}$ | 0.290 | 0.771 | 0.714 |
| $A_\pm$ & $B_\pm$ | $0.78^{+0.28}_{-0.26}$ | $11.8^{+5.3}_{-4.3}$ | $4.6^{+1.7}_{-1.1}$ | $3.3^{+1.4}_{-0.8}$ | $6.8^{+0.9}_{-0.8}$ | – | – | – |
| $C_+$ | $0.83^{+0.27}_{-0.27}$ | $6.0^{+2.1}_{-2.0}$ | $4.9^{+1.5}_{-1.1}$ | $3.7^{+0.8}_{-0.7}$ | $6.6^{+0.6}_{-0.6}$ | 0.757 | 0.034 | 0.083 |
| $C_-$ | $0.79^{+0.28}_{-0.26}$ | $5.8^{+2.1}_{-2.0}$ | $4.8^{+1.4}_{-1.0}$ | $3.6^{+0.7}_{-0.7}$ | $6.6^{+0.6}_{-0.6}$ | 0.938 | 0.041 | 0.123 |
| $D_+$ | $0.83^{+0.27}_{-0.28}$ | $6.1^{+2.0}_{-2.1}$ | $4.9^{+1.6}_{-1.1}$ | $3.8^{+0.8}_{-0.8}$ | $6.6^{+0.6}_{-0.6}$ | 0.625 | 0.014 | 0.028 |
| $D_-$ | $0.78^{+0.28}_{-0.26}$ | $5.8^{+2.1}_{-2.0}$ | $4.7^{+1.4}_{-1.0}$ | $3.7^{+0.7}_{-0.7}$ | $6.6^{+0.6}_{-0.6}$ | 1.000 | 0.012 | 0.039 |
| $C_\pm$ & $D_\pm$ | $0.81^{+0.28}_{-0.27}$ | $5.9^{+2.1}_{-2.0}$ | $4.8^{+1.5}_{-1.0}$ | $3.6^{+0.8}_{-0.7}$ | $6.6^{+0.6}_{-0.6}$ | – | – | – |
| All | $0.78^{+0.29}_{-0.26}$ | $11.2^{+5.5}_{-4.5}$ | $4.6^{+1.6}_{-1.1}$ | $3.5^{+1.3}_{-1.0}$ | $6.8^{+0.9}_{-0.8}$ | – | – | – |

**Table 9.** Physical parameters from Bayesian analysis for KMT-2021-BLG-1689

| Solution | Physical Properties | | | | | Relative Weights | | |
|---|---|---|---|---|---|---|---|---|
|  | $M_{\rm host}$ ($M_\odot$) | $M_{\rm planet}$ ($M_\oplus$) | $D_L$ (kpc) | $r_\perp$ (AU) | $\mu_{\rm rel}$ (mas/yr) | Gal. Mod. | $\chi^2$ | Total |
| A | $0.58^{+0.33}_{-0.27}$ | $46^{+30}_{-23}$ | $7.2^{+0.7}_{-1.2}$ | $2.2^{+0.4}_{-0.4}$ | $6.1^{+0.8}_{-0.7}$ | 0.993 | 1.000 | 1.000 |
| B | $0.57^{+0.33}_{-0.27}$ | $45^{+30}_{-22}$ | $7.2^{+0.7}_{-1.2}$ | $3.0^{+0.5}_{-0.6}$ | $6.1^{+0.8}_{-0.7}$ | 1.000 | 0.902 | 0.909 |
| A & B | $0.58^{+0.33}_{-0.27}$ | $46^{+30}_{-23}$ | $7.2^{+0.7}_{-1.2}$ | $2.5^{+0.8}_{-0.6}$ | $6.1^{+0.8}_{-0.7}$ | – | – | – |
| C | $0.68^{+0.40}_{-0.35}$ | $39^{+23}_{-20}$ | $5.0^{+1.5}_{-1.6}$ | $3.1^{+0.7}_{-0.9}$ | $10.9^{+1.7}_{-1.5}$ | 0.072 | 0.091 | 0.007 |
| D | $0.68^{+0.40}_{-0.35}$ | $39^{+24}_{-20}$ | $5.0^{+1.5}_{-1.6}$ | $3.5^{+0.8}_{-1.1}$ | $10.9^{+1.7}_{-1.5}$ | 0.068 | 0.139 | 0.010 |
| C & D | $0.68^{+0.40}_{-0.35}$ | $39^{+23}_{-20}$ | $5.0^{+1.5}_{-1.6}$ | $3.3^{+0.8}_{-1.0}$ | $10.9^{+1.7}_{-1.5}$ | – | – | – |
| All | $0.58^{+0.33}_{-0.27}$ | $46^{+30}_{-23}$ | $7.2^{+0.7}_{-1.3}$ | $2.5^{+0.8}_{-0.6}$ | $6.1^{+0.8}_{-0.7}$ | – | – | – |

NOTE. The $\mu_{\rm rel}$ values are different from those in Table 7, because extra prior (the Galactic model) is included.

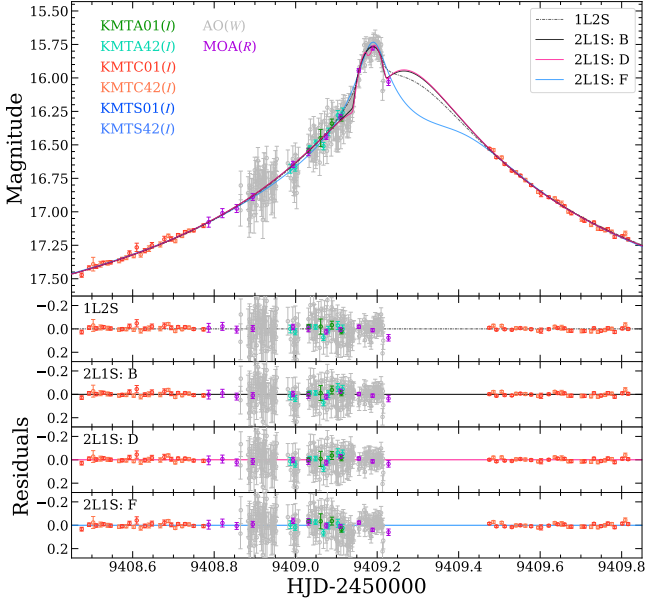

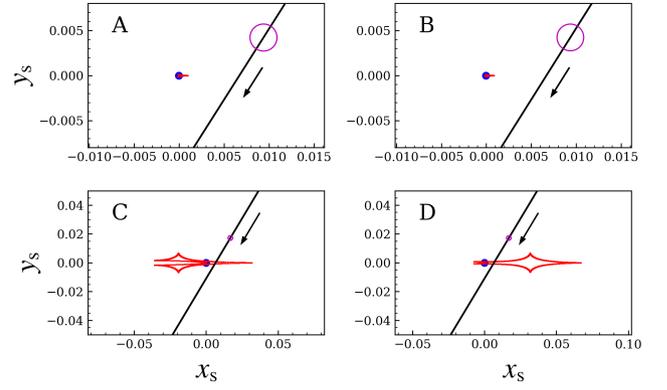

**Figure 6.** Caustic structures on the source plane for each binary-lens solution of KMT-2021-BLG-0171. Here $x_s$ and $y_s$ are in units of the angular Einstein radius $\theta_E$. The red lines represent the caustic position, and the blue dot is the location of the host star. The black line shows the source-lens relative trajectory, and the magenta circle represents the angular size of the source.

**Figure 4.** Light curve data of KMT-2021-BLG-1689 around the peak together with the best-fit models, 1L2S and 2L1S "B", "D" and "F". The 2L1S model names follow that in Fig. 8. The residuals for each model are shown in separate panels. For similar model pairs (A, B), (C, D), and (E, F) which do not show visible differences, we only plot one of each for clarity. The model light curve and data have been aligned to the KMTC $I$-band magnitude. The names and filters for each dataset are labeled on the panel, where $W$ means a Wratten #12 filter

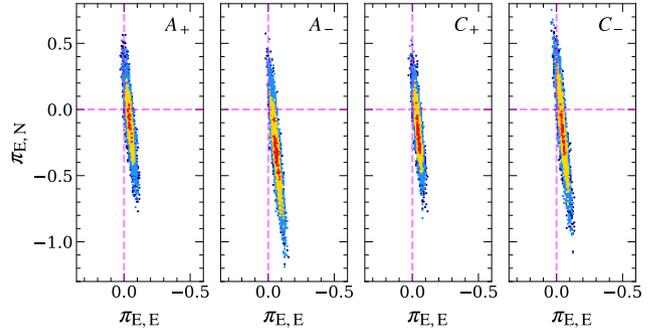

**Figure 7.** The $(\pi_{E,E}, \pi_{E,N})$ likelihood distribution of 2L1S Model $A_\pm$ and $C_\pm$ for KMT-2021-BLG-0171. The likelihood distribution of $(B_\pm, D_\pm)$ is nearly identical to that of $(A_\pm, C_\pm)$. The color (red, yellow, light blue, dark blue) indicates $[-2\ln(\mathcal{L}/\mathcal{L}_{\max})] < (1, 4, 9, \infty)$, respectively.

**Table 10.** Planetary sensitivity for KMT-2021-BLG-0171 with and without follow-up data ($-0.3 \leq \log s \leq 0.3$)

| log $q$ bin | KMT-2021-BLG-0171 | |
|---|---|---|
| | KMTNet only | KMTNet + follow-up |
| (−3.5, 3.0] | 0.9855 | 0.9991 |
| (−4.0, 3.5] | 0.7830 | 0.9338 |
| (−4.5, 4.0] | 0.4082 | 0.5984 |
| (−5.0, 4.5] | 0.1553 | 0.2882 |
| (−5.5, 5.0] | 0.0390 | 0.0888 |
| (−6.0, 5.5] | 0.0057 | 0.0106 |

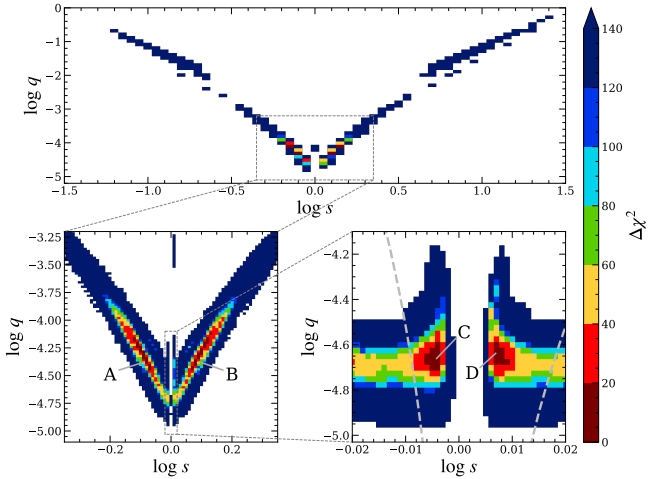

**Figure 5.** The projected $\chi^2$ distribution on the (log $s$, log $q$) plane from the grid search of KMT-2021-BLG-0171. The final solutions are labeled on the panels with their names. The upper panel shows the initial grid search, where (61 × 56) equally spaced grids were taken within the ranges of $-1.5 \leq \log s \leq 1.5$ and $-5.5 \leq \log q \leq 0$. The lower two panels are the refined grid searches near the local minima. The grid intervals of the lower left panel are $\Delta \log s = 0.01$ and $\Delta \log q = 0.02$, and the grid intervals of the lower right panel are $\Delta \log s = 0.001$ and $\Delta \log q = 0.025$. The grey dashed lines on the lower right panel represent the boundaries between central caustics and resonant caustics.

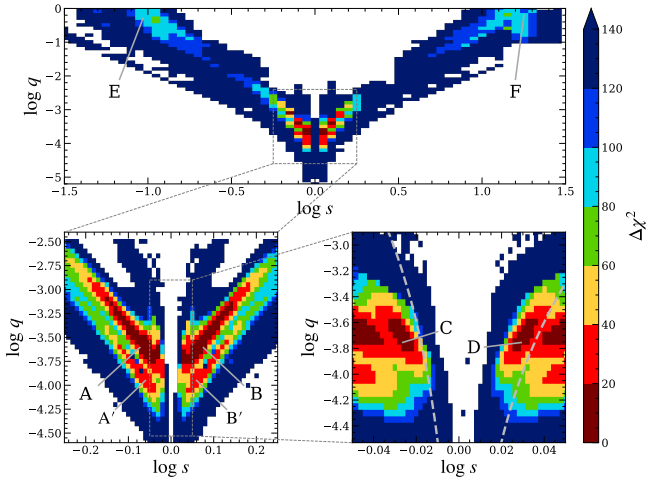

**Figure 8.** The projected $\chi^2$ distribution on (log $s$, log $q$) plane from the grid search of KMT-2021-BLG-1689. The final solutions are labeled on the panels with their names. The upper panel shows the initial grid search, where ($61 \times 56$) equally spaced grids were taken within the ranges of $-1.5 \leqslant \log s \leqslant 1.5$ and $-5.5 \leqslant \log q \leqslant 0$. The lower two panels are the refined grid searches near the local minima. The grid intervals of the lower left panel are $\Delta \log s = 0.01$ and $\Delta \log q = 0.05$, and the grid intervals of the lower right panel are $\Delta \log s = 0.002$ and $\Delta \log q = 0.05$. The grey dashed lines on the lower right panel represent the boundaries between central caustics and resonant caustics. The local minima (A′, B′) merge into (A, B) in the fitting if we allow log $\rho$ to vary.

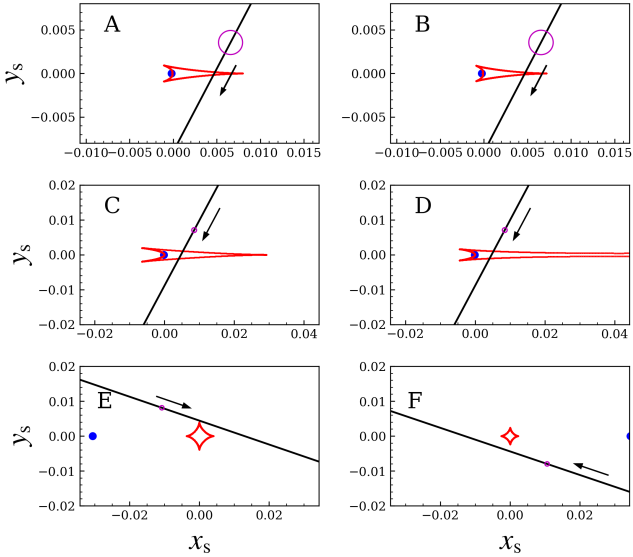

**Figure 9.** Caustic structures on the source plane for each binary-lens solution of KMT-2021-BLG-1689. Here $x_s$ and $y_s$ are in units of the angular Einstein radius $\theta_E$. In each panel, the red line represents the caustic position, and the blue dot is the location of the host star. The black line and arrow show the source-lens relative trajectory, and the magenta circle represents the source angular size (A, B, C, and D) or the $3\sigma$ upper limit of the source angular size (E and F). For A, C and E, the origin is set to the center of mass $x_{\mathrm{mass}}$. For B and D, the origin is set to the center of magnification of the primary lens, where $x_{\mathrm{mag},1} = x_{\mathrm{mass}} - (s - s^{-1})q/(1+q)$. For F, the origin is set to the magnification center of the secondary lens, where $x_{\mathrm{mag},2} = x_{\mathrm{mass}} + (s - s^{-1})/(1+q)$.

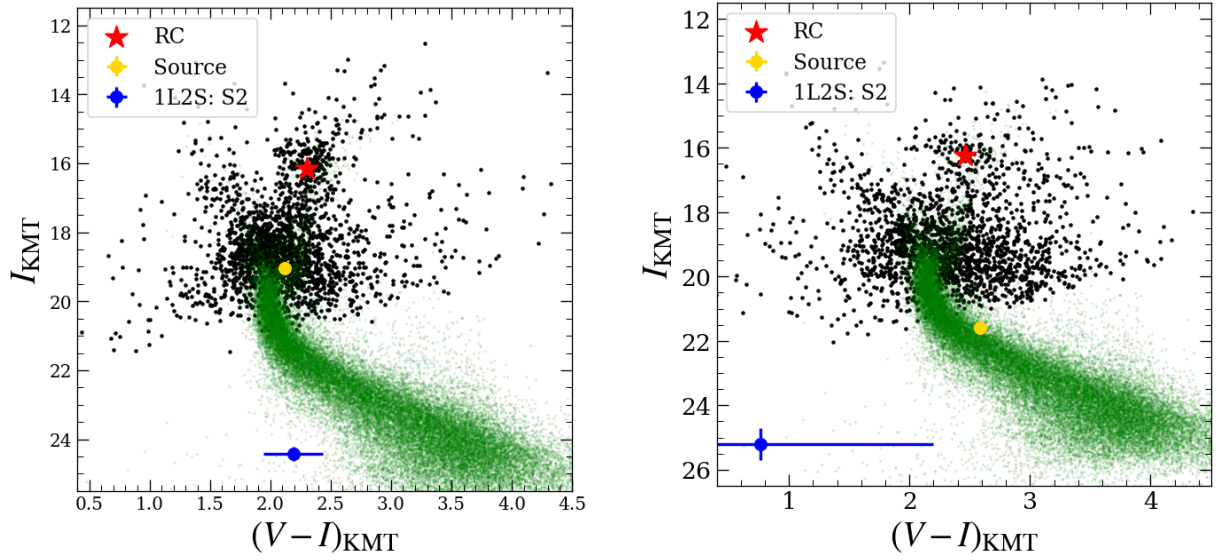

**Figure 10.** Color-magnitude diagrams (CMDs) for the $2' \times 2'$ square centered on each events. *Left panel*: KMT-2021-BLG-0171; *Right panel*: KMT-2021-BLG-1689. The black points are the field stars measured from KMTNet images. Green points are from the CMD obtained by Holtzman et al. (1998) from HST observations of Baade's Window, which we have aligned to the KMT CMD using the centroid of the red clump.. The positions of the red clump centroid (RC) and the microlens source are marked on the figure.

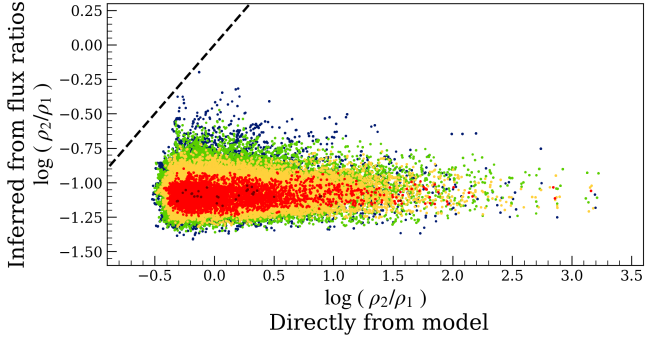

**Figure 11.** Source radius ratio of the 1L2S model for KMT-2021-BLG-0171 derived directly from the model values, $\rho_1$ and $\rho_2$, and inferred from the source flux ratios, $q_{f,V}$ and $q_{f,I}$. All the points are from the MCMC chain, and they are colored by $\Delta\chi^2 = \chi^2 - \chi^2_{\min,1L2S} < 1$ (dark red), $< 4$ (red), $< 9$ (yellow), $< 16$ (green), and $< \inf$ (blue). The dashed black line shows "$x$"="$y$". The distribution is considerably offset from the line, indicating that the 1L2S interpretation is not self-consistent.

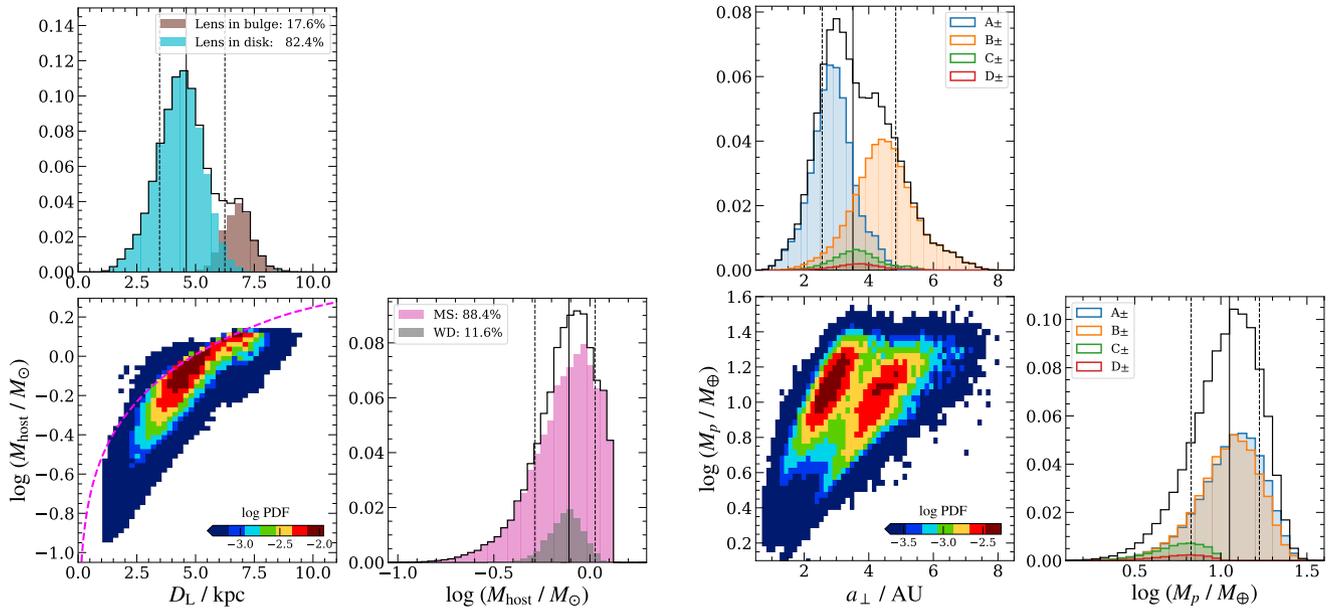

**Figure 12.** Bayesian posterior probability of the physical parameters of the lens system in KMT-2021-BLG-0171. The left panel shows the distribution of the lens system distance and the mass of the host star. The right panel shows the distribution of the planet mass and the projected separation to the host, different solutions are marked in different colors.

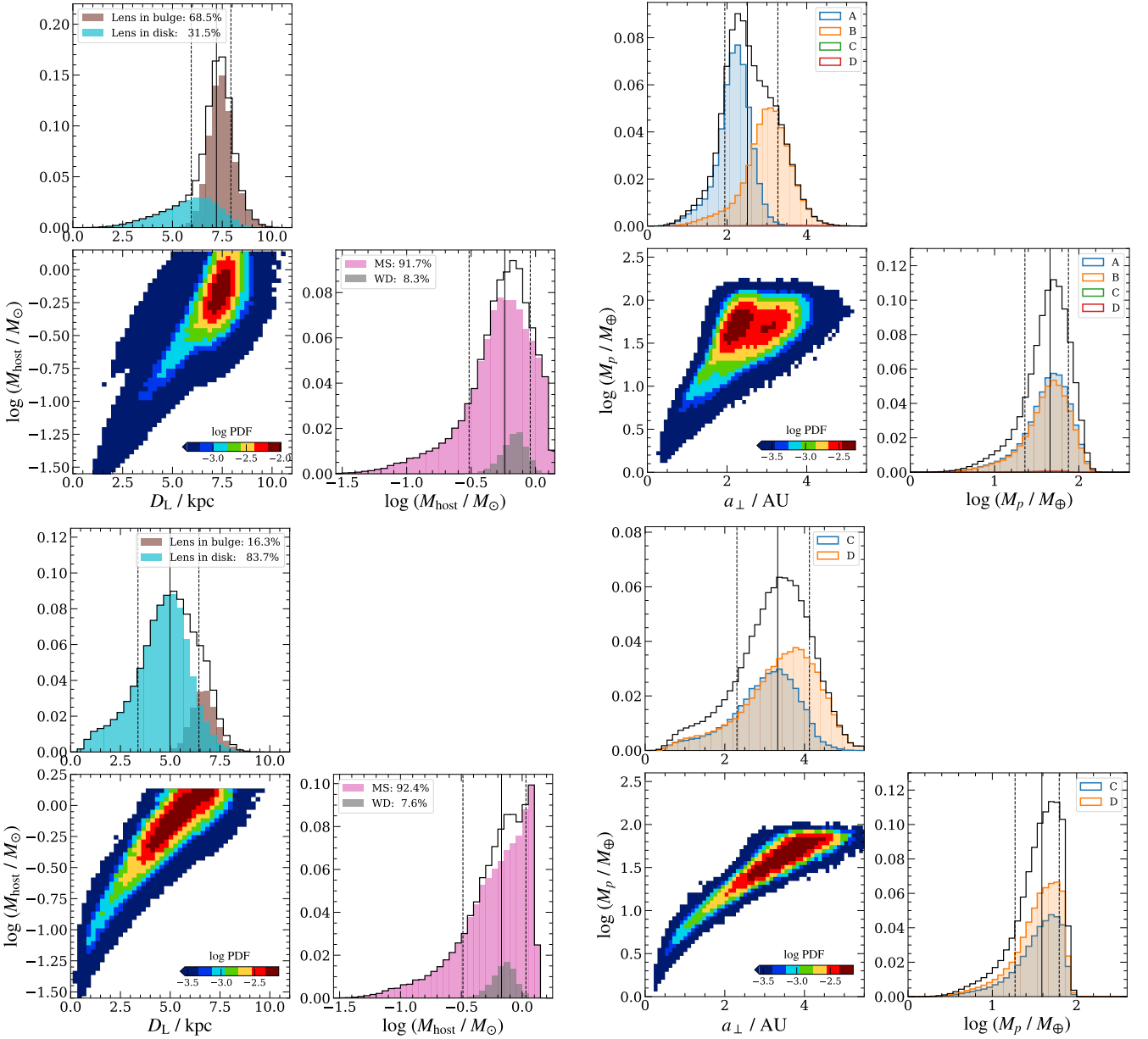

**Figure 13.** Bayesian posterior probability of the physical parameters of the lens system in KMT-2021-BLG-1689. *Top panels*: The combined distribution from all solutions which is dominated by solutions A and B. *Bottom panels*: The distributions for the (disfavored) C and D solutions alone. The notations of each panel are the same as in Fig. 12.

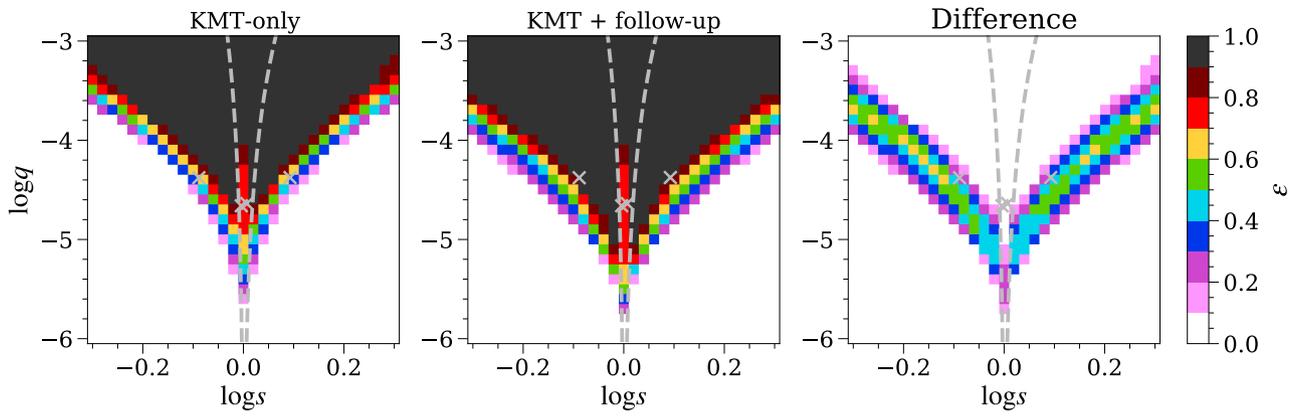

**Figure 14.** The planetary sensitivity as a function of $(\log s, \log q)$ for KMT-2021-BLG-0171 with and without follow-up data. The grey "×" marks are the actual planetary solutions. The grey dashed lines represent the boundaries between the separated and resonance caustic morphologies.

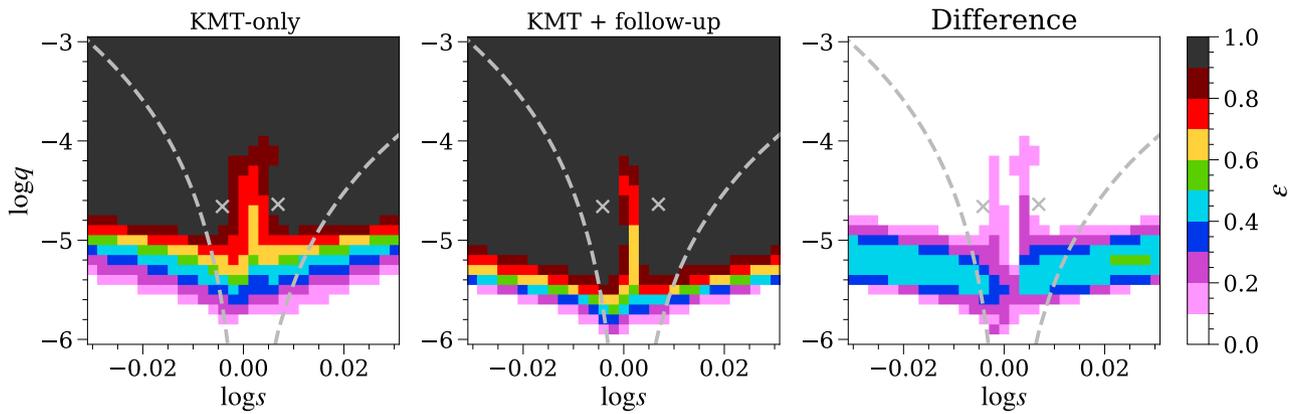

**Figure 15.** A zoom-in of Fig. 14 with denser $\log s$ sampling. The notations are the same as in Fig. 14.